\documentclass[aps,prx,twocolumn,superscriptaddress,noeprint,longbibliography, nofootinbib,floatfix]{revtex4-2}

\usepackage{graphicx}% Include figure files
\usepackage{bm}% bold math
\usepackage{amsmath, amssymb}
\usepackage{booktabs}
\usepackage{mathrsfs}
\usepackage[normalem]{ulem}
\usepackage{pifont}
\usepackage[mathscr]{euscript}
\usepackage{physics}
\usepackage{xcolor}
\usepackage{hyperref}

\newcommand{\quotes}[1]{``#1''}

\begin{document}

\title{Thermodynamics and fractal dynamics of nematic spin ice,\\ a doubly frustrated pyrochlore Ising magnet}

%%%%%%%%%%%%%%%%%%%%%%%%%%%%%%%%%%%%%%%%%%%%%%%%%%%%%%%%%%%%%%%%%
%%%%%%%%%%%%%%%%%%%%%%%%%%%%%%%%%%%%%%%%%%%%%%%%%%%%%%%%%%%%%%%%%
\author{Jonathan N. Hall\'en}
\address{TCM Group, Cavendish Laboratory, University of Cambridge, Cambridge CB3 0HE, UK}
\address{Max Planck Institute for the Physics of Complex Systems, 01187 Dresden, Germany}

\author{Claudio Castelnovo}
\address{TCM Group, Cavendish Laboratory, University of Cambridge, Cambridge CB3 0HE, UK}

\author{Roderich Moessner}
\address{Max Planck Institute for the Physics of Complex Systems, 01187 Dresden, Germany}

%%%%%%%%%%%%%%%%%%%%%%%%%%%%%%%%%%%%%%%%%%%%%%%%%%%%%%%%%%%%%%%%
%%%%%%%%%%%%%%%%%%%%%%%%%%%%%%%%%%%%%%%%%%%%%%%%%%%%%%%%%%%%%%%%

\date{\today}

%%%%%%%%%%%%%%%%%%%%%%%%%%%%%%%%%%%%%%%%%%%%%%%%%%%%%%%%%%%%%%%%
%%%%%%%%%%%%%%%%%%%%%%%%%%%%%%%%%%%%%%%%%%%%%%%%%%%%%%%%%%%%%%%%
\begin{abstract}
The Ising antiferromagnets on the triangular and on the pyrochlore lattices are two of the most iconic examples of magnetic frustration, paradigmatically illustrating many  exotic properties such as emergent gauge fields, fractionalisation, and topological order. In this work, we show that the two instances of frustration can, remarkably, be combined in a single system, where they coexist without inducing conventional long range ordering. 
Our results indicate that the system undergoes a first order phase transition upon lowering the temperature, into a yet different frustrated phase that we characterise to exhibit nematic order. We argue that an extensive degeneracy survives down to zero temperature, at odds with a customary Pauling estimate. Dynamically, we find evidence of anomalous noise in the power spectral density, arising from an effectively fractal anisotropic motion of monopoles at low temperature. 
\end{abstract}
%%%%%%%%%%%%%%%%%%%%%%%%%%%%%%%%%%%%%%%%%%%%%%%%%%%%%%%%%%%%%%%%
%%%%%%%%%%%%%%%%%%%%%%%%%%%%%%%%%%%%%%%%%%%%%%%%%%%%%%%%%%%%%%%%
\maketitle

%%%%%%%%%%%%%%%%%%%%%%%%%%%%%%%%%%%%%%%%%%%%%%%%%%%%%%%%%%%%%%%%
%%%%%%%%%%%%%%%%%%%%%%%%%%%%%%%%%%%%%%%%%%%%%%%%%%%%%%%%%%%%%%%%

\section{Introduction}
Frustration in materials -- occurring when competing interactions and lattice geometry lead to an exceptional abundance of states with similar energy and a suppression of conventional ordering -- has established itself as a rich context in which to study intriguing phenomena like topological order, fractionalised excitations, and unusual (typically slow) dynamics~\cite{lacroix2011}. 

In this paper we merge two of the most famous examples of frustrated magnetic systems: the venerable triangular lattice Ising antiferromagnet~\cite{wannier1950, Houtappel1950} and spin ice~\cite{udagawa2021spin}, arguably the best-established example of a Coulomb phase~\cite{PhysRevLett.91.167004,PhysRevB.69.064404,PhysRevLett.93.167204,henley2010}, topological order, and fractionalisation~\cite{Castelnovo_12}, with its characteristic emergent magnetic monopole excitations~\cite{Castelnovo2008}.

\begin{figure}
    \centering
    \includegraphics[width=\columnwidth]{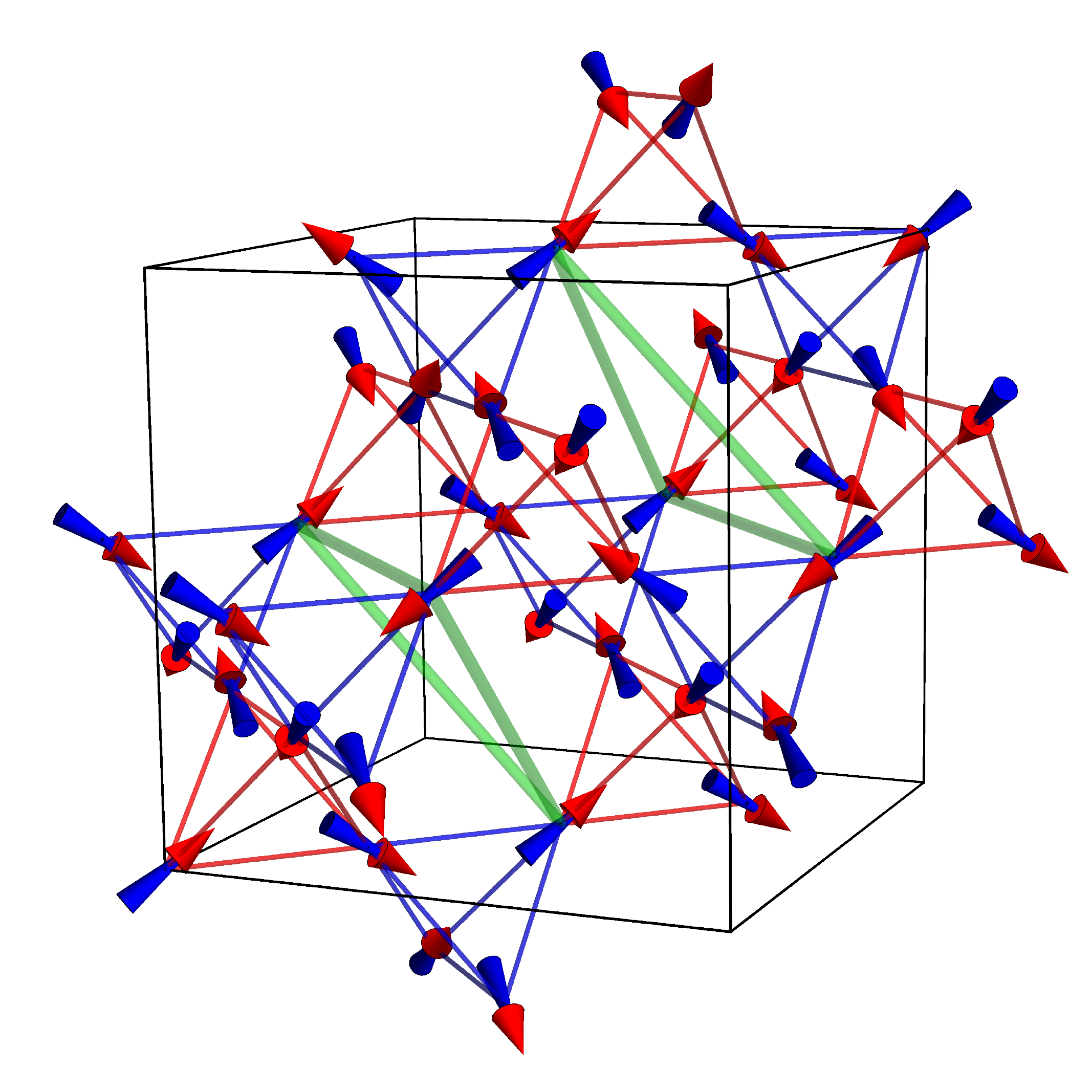}
    \caption{Spin ice consists of a pyrochlore lattice with classical Ising spins that are constrained to point directly in or out of a tetrahedron ($\sigma_i = \pm 1$ in Eq.~\eqref{eq:Hamiltonian}, with opposite convention on each of the two tetrahedral sublattices). The ferromagnetic nearest-neighbour interactions generate an extensive number of ground states with two spins pointing in and two spins pointing out of every tetrahedron. These \quotes{2-in 2-out} requirements are known as the ice rules. In our model, Eq.~\eqref{eq:Hamiltonian}, antiferromagnetic interactions between third-neighbour spins are also included in the direction perpendicular to the spin axis (conventionally referred to as $J_3'$ coupling). Examples of these interactions are illustrated by the thick green lines, which also illustrate how these interactions form triangular planes cutting through the pyrochlore lattice. To avoid cluttering, these interactions are only displayed for one of the four spin sublattices -- the others lie in the three symmetry-equivalent \{111\} planes. The black box indicates the size of a system with $L=1$ (i.e., the conventional 16-spin cubic unit cell of the pyrochlore lattice).}
    \label{fig:PSI+triangles}
\end{figure}

We focus on nearest-neighbour spin ice, defined on the pyrochlore lattice. The extensive set of its ground states is defined by configurations having two spins pointing into and two spins pointing out of each tetrahedron (see Fig.~\ref{fig:PSI+triangles}). 
%
%
%%%
%%%
%
%
Flipping a single spin in this state generates a monopole-antimonopole pair (3-out-1-in and 1-out-3-in tetrahedra). Creating the pair costs energy, but once created the monopoles can separate and move independently through further spin flips at zero energy cost. 

The very same set of spins on a pyrochlore lattice (which is tetrapartite) can be seen as four interpenetrating sets of 2D triangular layers, one set per sublattice. Nearest-neighbour interactions within each plane are one and the same as the so-called $J_3'$ third-neighbour interaction in spin ice~\cite{samarakoon2020}. Therefore, the physics of a spin system on the pyrochlore lattice endowed with only antiferromagnetic $J_3'$ interactions is akin to that of decoupled 2D triangular Ising antiferromagnetic layers, with their extensive degeneracy and characteristic `free spins' at the centres of minimal hexagonal loops with vanishing magnetisation~\cite{wannier1950}. 

Here we bring together both elements by studying nearest-neighbour spin ice ($J_1 \geq 0$) in presence of third-neighbour antiferromagnetic interactions ($J_3' \geq 0$)~\footnote{Note that the nearest-neighbour interaction is ferromagnetic in terms of the vector spins, but anti-ferromagnetic in terms of the pseudospin Ising variables used in Eq.~\eqref{eq:Hamiltonian} (see e.g., Ref.~\onlinecite{bramwell2020}). The third-neighbour interaction on the other hand has the same antiferromagnetic nature in both vector spin and pseudospin terms.}: 
%%%
\begin{equation}
    {\cal H} = J_1 \sum_{\langle i, j \rangle} \sigma_i \sigma_j + J_3' \sum_{\langle i, j \rangle_{3'}} \sigma_i \sigma_j
    \, .
    \label{eq:Hamiltonian}
\end{equation}
%%%
Surprisingly, we show that both energy terms can be concurrently minimised by extensively many ground states, leading to finite zero-point entropy, at odds with a na\"ive Pauling estimate predicting `negative entropy' and therefore conventional order (see App.~\ref{sec:pauling}). 
We study the thermodynamic properties and correlations in this system, to trace out its phase diagram and uncover the existence of a phase transition. To the best of our numerical understanding, the phase transition appears to be first order, separating a paramagnetic (spin ice) phase at high temperature from a low temperature nematic (spin ice) phase. Our results further confirm that the nematic phase remains extensively degenerate with a finite entropy down to zero temperature. 

At low temperatures, the response and equilibration properties of spin ice rely on the free motion of thermally activated defects~\cite{ryzhkin2005}, the magnetic monopoles~\cite{Castelnovo2008,jaubert2009}, which act as facilitators of magnetisation dynamics. Similarly, the free spins play a key role in the magnetisation dynamics of the triangular Ising antiferromagnet at low temperatures. Our model combines the two features and we find a rich dynamical interplay, with spin ice monopoles moving preferentially via flipping spins free of $J_3'$ energy cost.

In this work, we characterise how this affects the magnetisation dynamics in terms of anomalous behaviour of the corresponding magnetic noise. 
In particular, we identify anisotropic fractal structures on which monopoles move at low temperatures, which we show to be responsible for the observed anomalous exponents. This is notably similar, albeit of a remarkably different origin, to the behaviour recently observed in Ref.~\onlinecite{hallen2022}. 

While we are able in the end to paint a reasonably complete picture of the thermodynamics and dynamic properties of this model, some intriguing open questions remain for future work -- from the exact counting of the nematic ground states, to the fate of the first order transition in the limit of $J_3' \gg J_1$.

In Sec.~\ref{sec:NSIstates} we describe the extensive set of exact ground states for ${\cal H}$, and explain how these break rotational symmetry. Following this, we turn to the thermodynamic behaviour of the model in Sec.~\ref{sec:ThermoDynamics}, where we use Monte Carlo simulations to characterise the nematic order further and to study the nature of the phase transition between conventional and nematic spin ice. In Sec.~\ref{sec:PSD} magnetic noise measurements from Monte Carlo simulations are presented, displaying decay with an anomalous exponent when the system is in the nematic phase. The anomalous exponent is explained in Sec.~\ref{sec:Clusters}, where we show that the clusters of free spins -- on which the monopoles are biased to move -- are fractal on the relevant length scales. We conclude by commenting on the importance of these results, and by highlighting a number of questions that remain open for further work.
In the conclusions we also comment on the relevance of our results to recent experiments on anomalous noise and farther range exchange terms in realistic spin ice Hamiltonians. 

The manuscript includes a number of appendices where selected details are conveniently presented in a way that does not interrupt the flow of the main narrative: Pauling's estimate (App.~\ref{sec:pauling}); lattice commensurability (App.~\ref{app:SystemSize}); numerical methods (App.~\ref{app:Methods}); ground state construction (App.~\ref{app:GSbuilding}) and local correlators (App.~\ref{app:correlators}); supercooling limits (App.~\ref{app:Supercooling}). 
%
%
%%%%%%%%%%%%%%%%%%%%%%%%%%%%%%%%%%%%%%%%%%%%%%%%%%%%%%%%%%
%
%
%%%
\begin{figure}[ht!]
    \centering
    \includegraphics[width=\columnwidth]{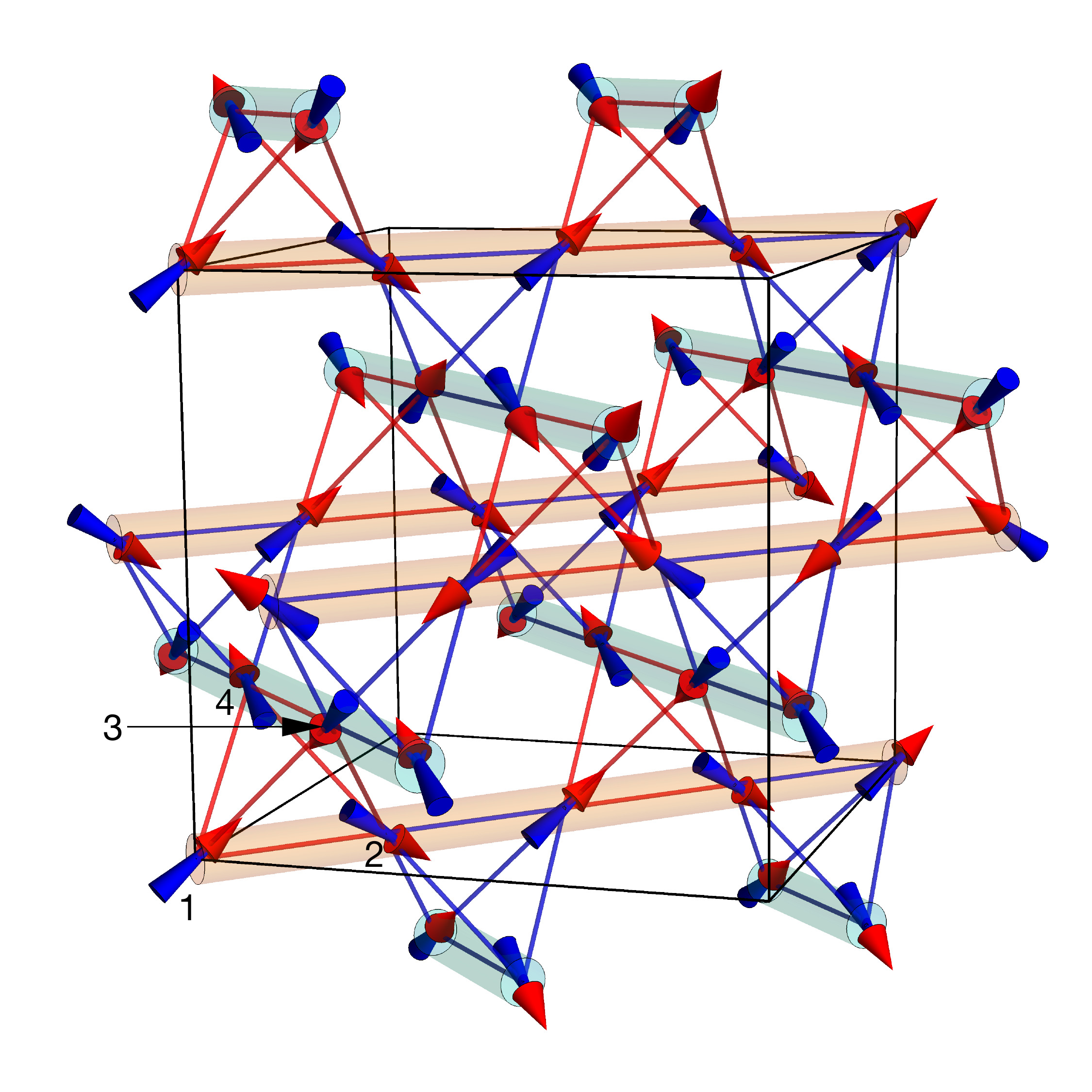}
    \caption{Illustration of the $\beta_{12}$ (orange) and and $\beta_{34}$ (cyan) spin chains, pointing along the [110] and [-110] directions, respectively. The numbers in black indicate the four spin sublattices. The black box indicates the size of a system with $L=1$ (i.e., the conventional 16-spin cubic unit cell of the pyrochlore lattice). Every spin belongs to one of these $\beta$ chains, and any state where all spins along each chain are aligned satisfies the ice rules. In addition, if the direction of these chains of aligned spins is chosen correctly (see main text), the energy of the third-neighbour interactions will also be minimised, generating a ground state of ${\cal H}$. Two alternative pairs of $\beta$-chains can be defined in the same way, $\beta_{13}$ and $\beta_{24}$ or $\beta_{14}$ and $\beta_{23}$.}
    \label{fig:Chains}
\end{figure}
%%%
%
%
%%%%%%%%%%%%%%%%%%%%%%%%%%%%%%%%%%%%%%%%%%%%%%%%%%%%%%%%%%

\section{Nematic spin ice states
\label{sec:NSIstates}}
The third-neighbour interactions form triangular lattice planes, cutting through the pyrochlore lattice with four different $\{111\}$ orientations. 
The surprising discovery here is that an extensive number of spin ice states are also ground states of all the triangular planes. These states break rotational symmetry without the appearance of long-range translational order, hence the name nematic spin ice. The nematic order resembles that observed in Ref.~\onlinecite{chern2008} for a model of Heisenberg spins on the pyrochlore lattice. There are, however, several important differences, including the fact that the nematic order in our model appears to persist to zero temperature. The rotational symmetry breaking is most clearly demonstrated by the behaviour of the spin-spin correlations and the order parameter $\Gamma$ (see Sec.~\ref{sec:ThermoDynamics}).

The nematic spin ice phase is characterised by the selection of one pair of $\beta$-chains in spin ice (see Fig.~\ref{fig:Chains}), along which spins preferably align head to tail. There are three different possible choices of chain pairs  (normal to the three global crystallographic axes), corresponding to the three possible ways of combining the four sublattices. Throughout this paper we choose our coordinate system such that the selected chain pair is $\beta_{12}$-$\beta_{34}$ in all realisations.

Every spin in a tetrahedron belongs to a different triangular plane, and for an $L\times L \times L$ system with a $16$ spin unit cell and periodic boundary conditions there are $4L$ such planes. In order for the tripartite sublattice structure of the triangular lattice planes to be well defined in a system with periodic boundary conditions, $L$ must be an integer multiple of $3$ (see App.~\ref{app:SystemSize}). Antiferromagneticically interacting Ising spins on a triangular lattice was one of the earliest frustrated models to be studied~\cite{wannier1950, Houtappel1950}, and it supports an extensive number of degenerate ground states. In these ground states, with energy $-J_3'$ per spin, every triangle has either two spins up and one down or one spin up and two down. Excitations take the form of three-up or three-down triangles, and each of these cost $4 J_3'$ to create.

In App.~\ref{app:GSbuilding}, we show how a subextensive number of ground states for ${\cal H}$ in Eq.~\eqref{eq:Hamiltonian} can be constructed exactly, with all spins aligned along the selected chains.
Importantly, ground states of the triangular lattice Ising antiferromagnet typically have large numbers of spins which are free to flip without energy cost (namely, spins that sit at the centre of minimal hexagonal rings on the lattice with vanishing magnetisation). These free spins form channels through which magnetic monopoles can move without incurring any energy cost. By creating a  pair of monopoles, moving them in a closed loop consisting of free spins, and then annihilating them, it is possible to move between different ground states of the system. These loops bring us from the subextensive degeneracy of the fully aligned chain states described above to the extensive degeneracy of the full set of ground states. Heuristically, we observe that such free loops exist in a general ground state found through simulated annealing from high temperature. In App.~\ref{app:GSbuilding} a rigorous argument is presented, showing that there must be a finite density of local, free loops.~\footnote{Note that, for a system with only short-ranged interactions, the existence of a finite density of local (non-winding) updates that bring the system between ground states guarantees that the number of ground states grows exponentially with system volume.}

%%%%%%%%%%%%%%%%%%%%%%%%%%%%%%%%%%%%%%%%%%%%%%%%%%%%%%%%%%%
%%%%%%%%%%%%%%%%%%%%%%%%%%%%%%%%%%%%%%%%%%%%%%%%%%%%%%%%%%%

\section{Thermodynamic behaviour \label{sec:ThermoDynamics}}
\begin{figure}
    \centering
    \includegraphics[width=\columnwidth]{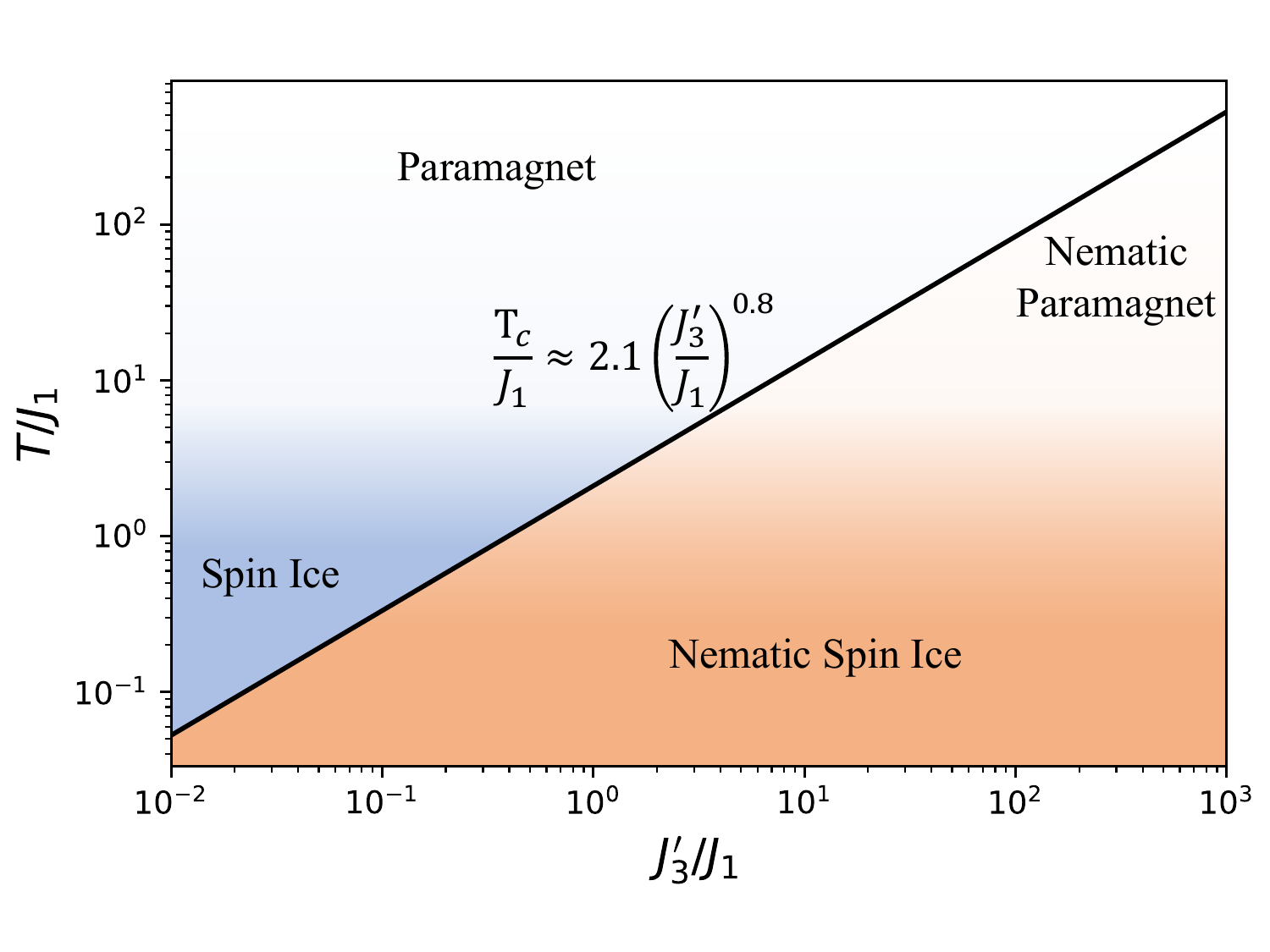}
    \caption{Illustrative phase diagram of the Hamiltonian in Eq.~\eqref{eq:Hamiltonian}. Both the high temperature (cooperative) paramagnetic phase, shown in blue, and the low temperature spin nematic phase, shown in red, have a spin ice regime at low temperature. The phases are separated by a phase transition at $T_c/J_1 \approx 2.1 (J_3'/J_1)^{0.8}$, with the critical temperature determined in Fig.~\ref{fig:orderpar} and \ref{fig:SpecHeat}.}
    \label{fig:PhaseDiagram}
\end{figure}

In this section we discuss the thermodynamic equilibrium behaviour of ${\cal H}$, analysing the four regimes noted in the phase diagram in Fig.~\ref{fig:PhaseDiagram}, and characterising the transition between them. 
%
%
%%%
%%%
%
%

\begin{figure}
    \centering
    \includegraphics[width=\columnwidth]{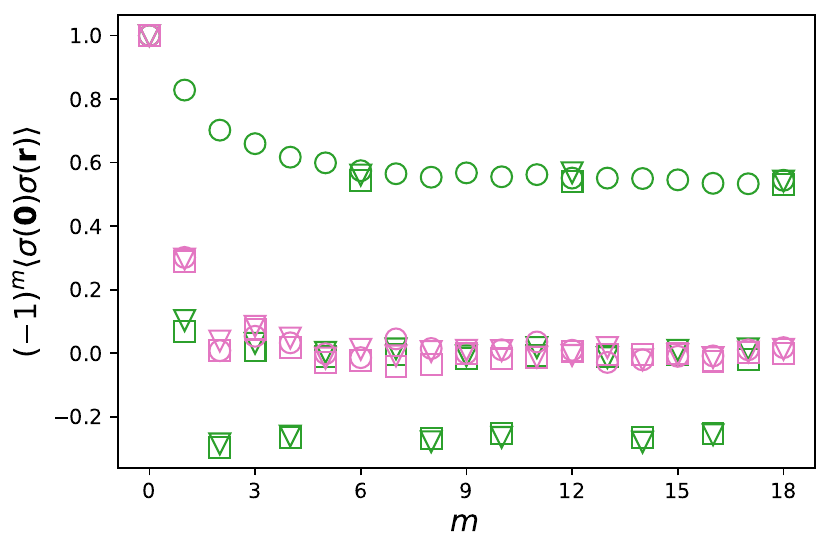}
    \caption{Spin correlations along the three chain directions $\beta_{12}$ (circles),  $\beta_{13}$ (squares), and  $\beta_{14}$ (triangles) for $J_3'/J_1=0.2$ at temperatures $T/J_1=0.2 < T_c/J_1$ (green) and $T/J_1 = 0.9 > T_c/J_1$ (pink). The three directions correspond to $\mathbf{r}=(m, m,0)$, $\mathbf{r}=(m, 0,m)$, and $\mathbf{r}=(0, m,m)$ respectively, with integer $m$. Below $T_c$ one chain direction ($\beta_{12}$ in this case) is chosen along which spins tend to align head to tail, resulting in large spin correlation along this direction. The staggered behaviour of the spin correlations in the other directions, with non-zero correlations for even $m$, is a result of the antiferromagnetic interactions between chains of the same type (see App.~\ref{app:correlators} for details).
    }
    \label{fig:spinspincorr}
\end{figure}

That the nematic state corresponds to the alignment of spins along one spontaneously selected pair of $\beta$ chains can be observed by considering the spin correlators $(-1)^m \langle \sigma(\mathbf{0}) \sigma(\mathbf{r}) \rangle$ along the chains, where $m$ is an integer equal to the number of steps between $\mathbf{0}$ and  $\mathbf{r}$. (Alignment of the vector spins along a chain corresponds to Ising variables that follow the pattern $+,-,+,-,...$ -- the factor of $(-1)^m$ is included to account for this.) Above the transition temperature, the correlations are short-ranged and there is no difference between the different chain directions. Below the phase transition, long range correlations appear along the chains and one set of directions is chosen along which spins preferentially align head-to-tail (see Fig.~\ref{fig:spinspincorr} and the further discussion in App.~\ref{app:correlators}).
%
%
%%%
%%%
%
%

To analyse the nematic order, we define a Potts variable $\gamma_t$ on each tetrahedron in the system
%%%
\begin{equation}
    \gamma_t = 
    \begin{cases}
     {\rm e}^{i\frac{\pi}{3}} & \text{if $\sigma_{t,1}=\sigma_{t,3}=-\sigma_{t,2}=-\sigma_{t,4}$} \\
     -1 & \text{if $\sigma_{t,1}=\sigma_{t,2}=-\sigma_{t,3}=-\sigma_{t,4}$} \\
     {\rm e}^{-i\frac{\pi}{3}} & \text{if $\sigma_{t,1}=\sigma_{t,4}=-\sigma_{t,2}=-\sigma_{t,3}$} \\
     0 & \text{otherwise} 
    \end{cases}
    \label{eq:PottsVar}
    \end{equation}
%%%
where $\sigma_{t, a}$ is the $a$ sublattice spin ($a=1,2,3,4$; see Fig.~\ref{fig:Chains}) in the tetrahedron indexed by $t$. $\gamma_t$ is zero if 
the spins on the tetrahedron do not obey the ice rules. 
We define $\Gamma$ as the average of $\gamma_t$ over all the $N_t$ tetrahedra in the system
%%%
\begin{equation}
    \Gamma = \frac{1}{N_t}\sum_{t=1}^{N_t} \gamma_t
    \, .
    \label{eq:GammaDef}
\end{equation}
%%%
In a typical configuration of normal spin ice the three non-zero values of $\gamma_t$ are equally distributed, and therefore $\langle \Gamma \rangle =0$ (where $\langle ... \rangle$ indicates a thermodynamic average). However, in the nematic phase one pair of $\beta$-chains is spontaneously selected along which spins tend to align. The result is that two of the three non-zero values of $\gamma_t$ are favoured as compared to the third one in the nematic phase, and $\langle \Gamma \rangle$ becomes non-zero with complex phase $0$ if the selected chain pair is $\beta_{12}$-$\beta_{34}$, $-2\pi/3$ for $\beta_{13}$-$\beta_{24}$, and $+2\pi/3$ for $\beta_{14}$-$\beta_{23}$. Note that spins can align in two directions along the chains and two chains of the selected types cut through each tetrahedron. Consider, for example, a ground state formed by all spins aligned along the $\beta_{12}$-$\beta_{34}$ chains, such as the one shown in Fig.~\ref{fig:Chains}. From Eq.~\eqref{eq:PottsVar} it follows that a given tetrahedron in such a state will have $\gamma_t={\rm e}^{-i\frac{\pi}{3}}$ or $\gamma_t={\rm e}^{+i\frac{\pi}{3}}$ with equal probability. 

\begin{figure}[ht!]
    \centering
    \includegraphics[width=\columnwidth]{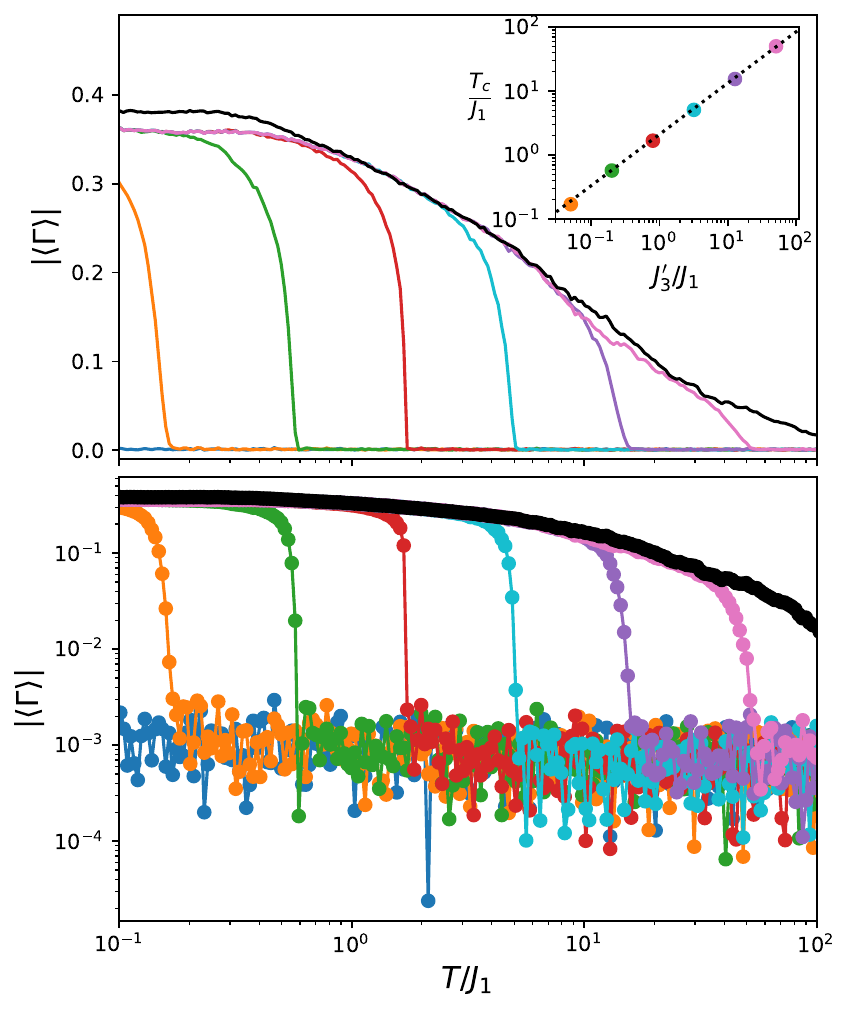}
    \caption{Behaviour of the magnitude of $\langle \Gamma \rangle$, defined in Eq.~\eqref{eq:GammaDef}. 
    Going from left to right, the value of $J_3'$ increases by factors of 4 between $J_3'/J_1=0.05$ (orange) and $J_3'/J_1=51.2$ (pink). The blue line corresponds to $J_3'=0$ (i.e., conventional nearest-neighbour spin ice, where $\Gamma$ vanishes at all temperatures). The black line corresponds to infinite $J_3'$. $\langle \Gamma \rangle$ acts as an order parameter, remaining zero up to the point where the system enters the nematic spin ice phase below the $J_3'$-dependent critical temperature $T_c$. We define pragmatically $T_c$ as the highest temperature where $|\langle \Gamma \rangle | > 10^{-2}$, and we plot it as a function of $J_3'$ in the inset (notice the logarithmic scale on both axes). The dotted black line $T_c/J_1=2.1 (J_3'/J_1)^{0.8}$ is a fit to the data.}
    \label{fig:orderpar}
\end{figure}

The strength and type of nematic order is captured by the magnitude and phase of $\langle \Gamma \rangle$, and it thus acts as an order parameter~\footnote{While we are unable to prove the absence of any underlying magnetic order, we were able to find only evidence for nematic order in our system, and therefore we adopted the corresponding nomenclature in the manuscript.}. 
Fig.~\ref{fig:orderpar} shows the behaviour of the magnitude of $\langle \Gamma \rangle$ as the system is cooled into the nematic spin ice phase. 
%
%
%%%
%%%
%
%
We observe a sudden increase in $|\langle \Gamma \rangle |$ consistent with a phase transition occurring at some temperature $T_c$ that depends on the parameters of the system. Specifically, $T_c$ increases with increasing $J_3'$ and it appears to diverge in the $J_3' \to \infty$ limit. Further information about the nematic phase is provided by considering correlators of the type $\langle \gamma_t \gamma_{t'} \rangle$, as described in App.~\ref{app:correlators}.

\begin{figure}
    \centering
    \includegraphics[width=\columnwidth]{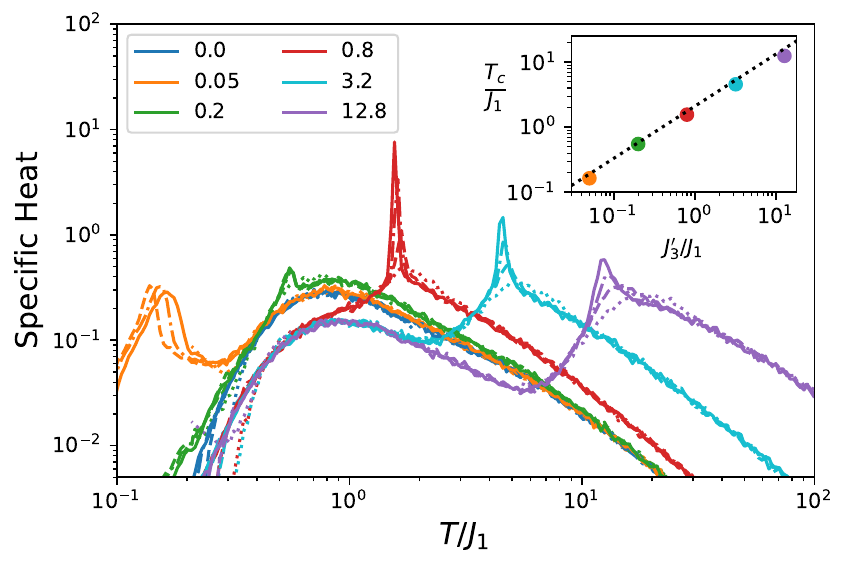}
    \caption{Specific heat per spin for ${\cal H}$ with different values of $J_3'/J_1$ (shown in the legend), computed for different system sizes: $L=3$ (dotted lines), $L=6$ (dashed lines), $L=9$ (dash-dotted lines), and $L=12$ (solid lines). The phase transition manifests itself as a peak in the specific heat, and the peak position depends only weakly on system size. The inset shows the critical temperature $T_c$, defined by the position of the specific heat peak for $L=12$, for each of the $J_3'$ values. The black dotted line shows $T_c/J_1=2.1 (J_3'/J_1)^{0.8}$, demonstrating the good agreement with the behaviour inferred from the order parameter in Fig.~\ref{fig:orderpar}.}
    \label{fig:SpecHeat}
\end{figure}

In the thermodynamic limit, the critical temperature $T_c(J_3')$ is the point where $\langle \Gamma \rangle$ becomes non-zero. For a finite-size system with $L=9$, $|\langle \Gamma \rangle |$ fluctuates with fluctuations of typical size $\sim 10^{-3}$ in the spin ice phase ($J_3'=0$) (see Fig.~\ref{fig:orderpar}), and we pragmatically define $T_c$ as the largest temperature where $|\langle \Gamma \rangle | > 10^{-2}$ for a given $J_3'$. Using this definition, we find that the behaviour of the critical temperature is well reproduced by $T_c/J_1 \propto (J_3'/J_1)^{0.8}$ (see inset of Fig.~\ref{fig:orderpar}). The exponent $0.8$ is determined from a fit, and at this point in time does not have an analytical explanation. We observe the same dependence on $J_3'$ if we instead extract $T_c$ from the position of the peak in the specific heat  (see Fig.~\ref{fig:SpecHeat}). 
%
%
%%%
%%%
%
%
We have no indication that this behaviour does not persist for $J_3' \to \infty$, and simulations with effectively infinite $J_3'$ (i.e., when the nearest-neighbour interaction $J_1$ is introduced in the extensively degenerate ensemble of configurations that are ground states of the triangular Ising antiferromagnet) have shown that the nematic phase persists up to at least $T/J_1=10^3$.

The phase transition related to rotational symmetry breaking appears to persist up to infinite $J_3'$, but the transition becomes progressively softer (see Fig.~\ref{fig:orderpar}). For larger $J_3'$ the transition occurs at higher temperatures, where the ice rules are less satisfied (see Fig.~\ref{fig:Densities}) and the transition is essentially between a paramagnetic state, with $\Gamma = 0$, and a nematic paramagnetic state below $T_c$, with a small but non-zero $\Gamma$. 
For smaller $J_3'$ the transition occurs instead at lower temperature, where the nearest-neighbour interactions play a more significant role, forcing the system closer to a state that satisfies the ice rules. This has the effect of increasing the magnitude of $\Gamma$. The phase transition then takes place between a generic spin ice state and a nematic spin ice state, and becomes sharper. 
%
%
%---------------------------------------------------------

\subsection{Nature of the phase transition \label{app:PhaseTransition}}
The behaviour of the order parameter at the transition is abrupt, as one can see clearly in Fig.~\ref{fig:orderpar}, and it is suggestive of first order behaviour. 

\begin{figure*}[ht!]
    \centering
    \includegraphics[height=0.9\textwidth]{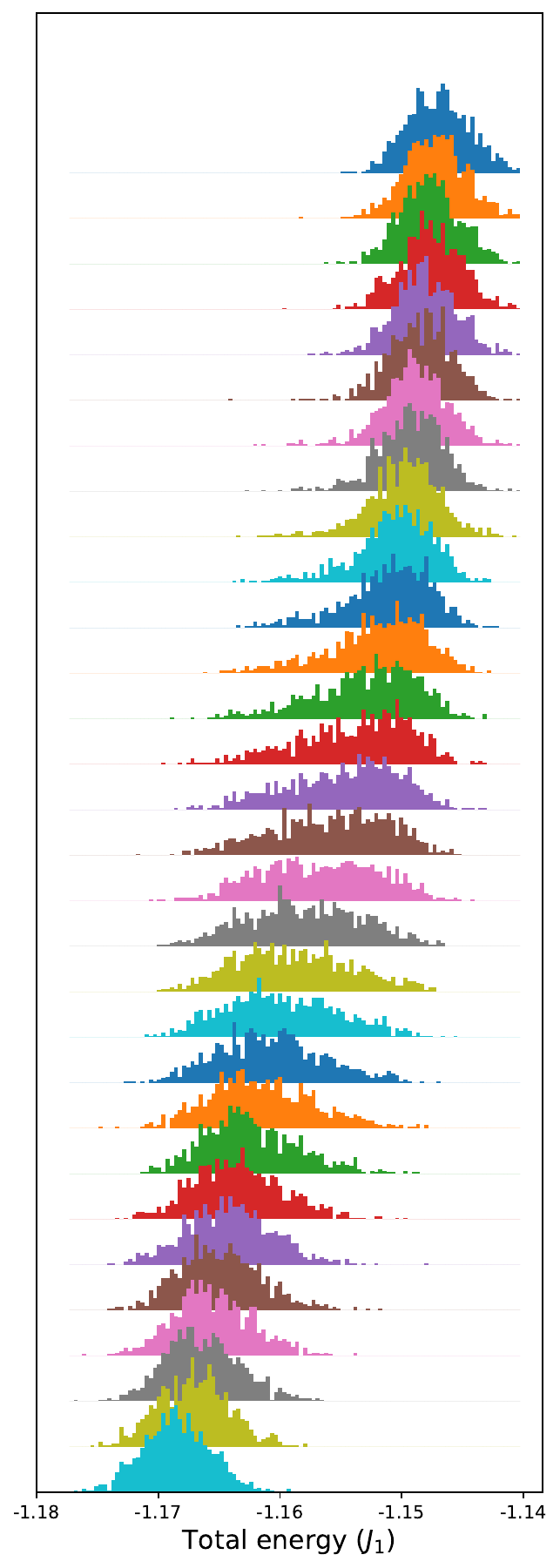}
    \includegraphics[height=0.9\textwidth]{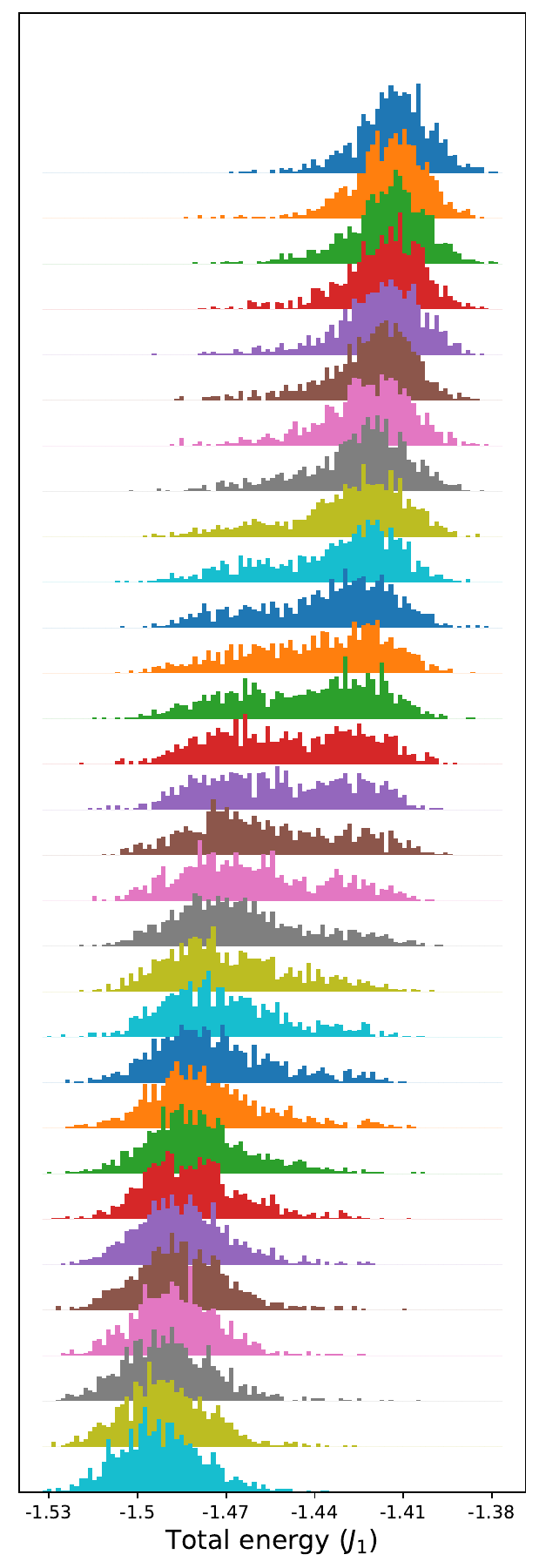}
    \includegraphics[height=0.9\textwidth]{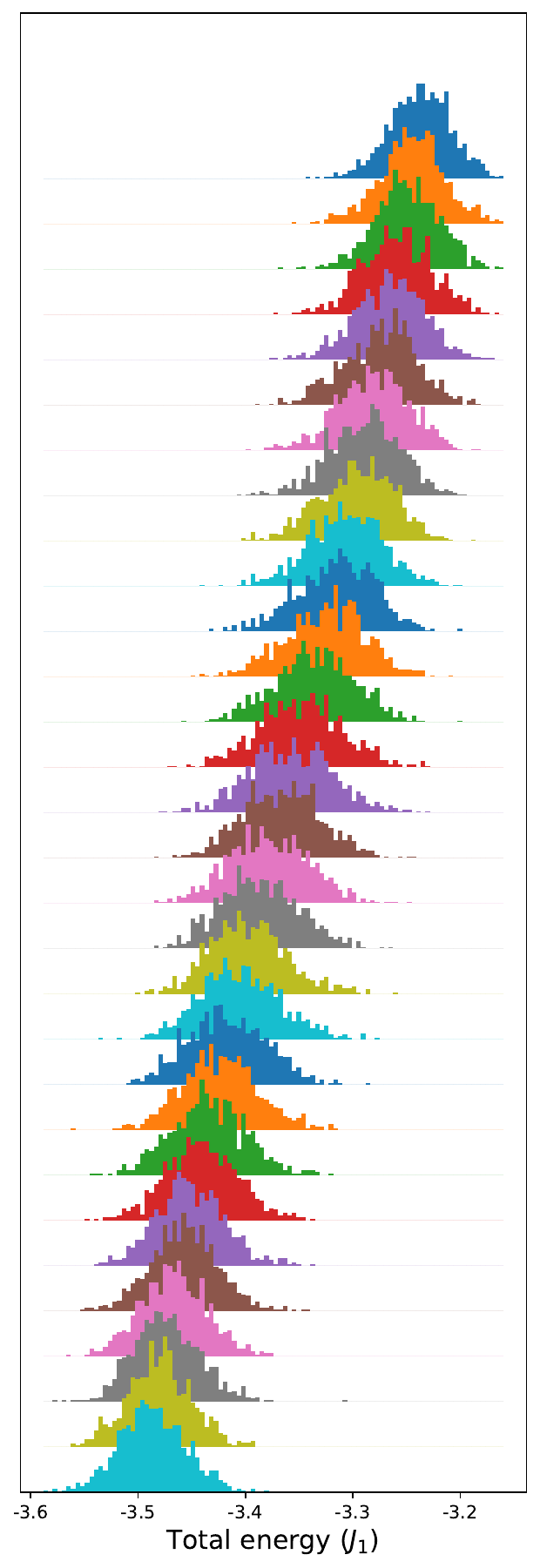}
    \caption{Histograms of the system energy at temperatures around the phase transition, for $J_3'/J_1=0.2$ (left), $J_3'/J_1=0.8$ (centre), and $J_3'/J_1=3.2$ (right). The data were collected from $1000$ independent simulations with $L=9$. From bottom to top (the curves are shifted for visualisation purposes), the temperature is varied from $0.506J_1$ to $0.538J_1$ for $J_3'/J_1=0.2$, from $1.650J_1$ to $1.698J_1$ for $J_3'/J_1=0.8$, and from $4.578J_1$ to $4.994J_1$ for $J_3'/J_1=3.2$. The temperature ranges were chosen to encompass the specific heat peak in each case, and the sample temperatures are distributed evenly in $\log{T}$ within each range.}
    \label{fig:Histograms}
\end{figure*}

To investigate the nature of the phase transition in greater detail, we considered histograms of the system energy near $T_c$ (Fig.~\ref{fig:Histograms}). For $J_3' / J_1 \simeq 0.8$, where the onset of the order parameter appears to be most sudden (Fig.~\ref{fig:orderpar}), we observe two characteristic peaks in the probability distribution function; as the temperate decreases across the transition the peak at higher energy decreases in intensity while the peak at lower energy grows. 
This is consistent with a first order transition.
For smaller values of $J_3'/J_1$, the behaviour becomes less pronounced, although it remains vaguely recognisable. For larger values of $J_3'/J_1$, we are actually unable to see a two-peak structure (at least within the resolution afforded by our finite size numerical simulations), and we observe instead a single peak which shifts smoothly to lower energy as the temperature is decreased. Examples are shown in Fig.~\ref{fig:Histograms}.
%
%
%%%
%\onecolumngrid
%\begin{center}
%\end{center}
%\twocolumngrid
%%%
%
%

We also looked at the behaviour of the Binder cumulants~\cite{binder1981} for the real and imaginary parts (denoted $\Gamma'$ and $\Gamma''$, respectively) of the order parameter $\Gamma$ in Eq.~\eqref{eq:GammaDef}, defined as~\footnote{Note that to correctly compute $V_\Gamma$ one must average over the three possible choices of directions for the order parameter -- i.e., in this case we do not rotate our coordinate axes so that each individual realisation favours alignment along the $\beta_{12}$ and $\beta_{34}$ chains.}: 
%%%
\begin{equation}
    V_{\Gamma} = 1 - \frac{\left\langle \left(\Gamma'\right)^4\right\rangle }{3 \left\langle \left(\Gamma'\right)^2\right\rangle^2}
\, , 
\end{equation}
%%%
and equivalently for the imaginary part. (The real and imaginary parts of $\Gamma$ are statistically equivalent.) 
%
%
%%%
\begin{figure}
    \centering
    \includegraphics[width=\columnwidth]{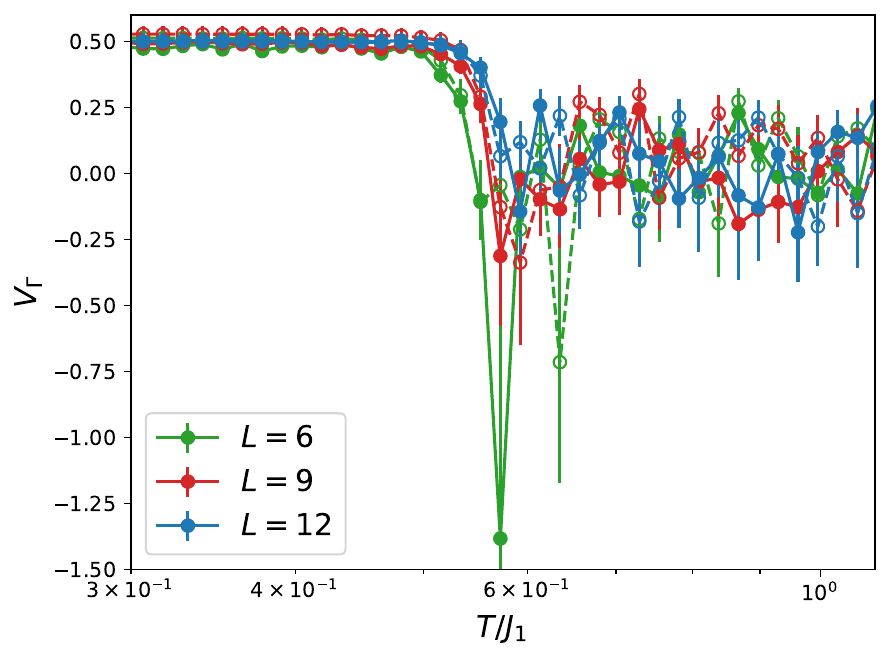}
    \includegraphics[width=\columnwidth]{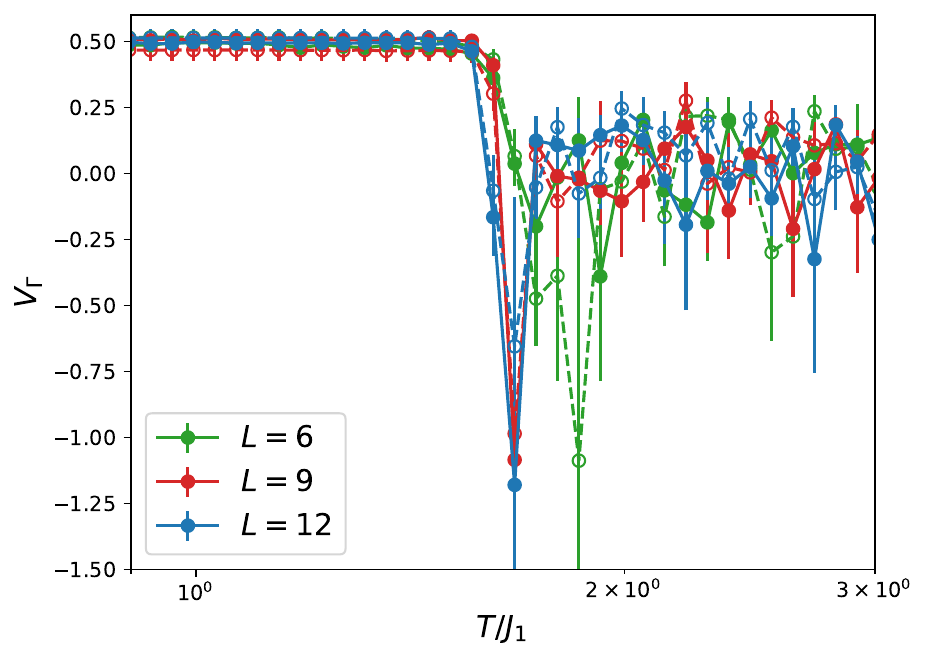}
    \includegraphics[width=\columnwidth]{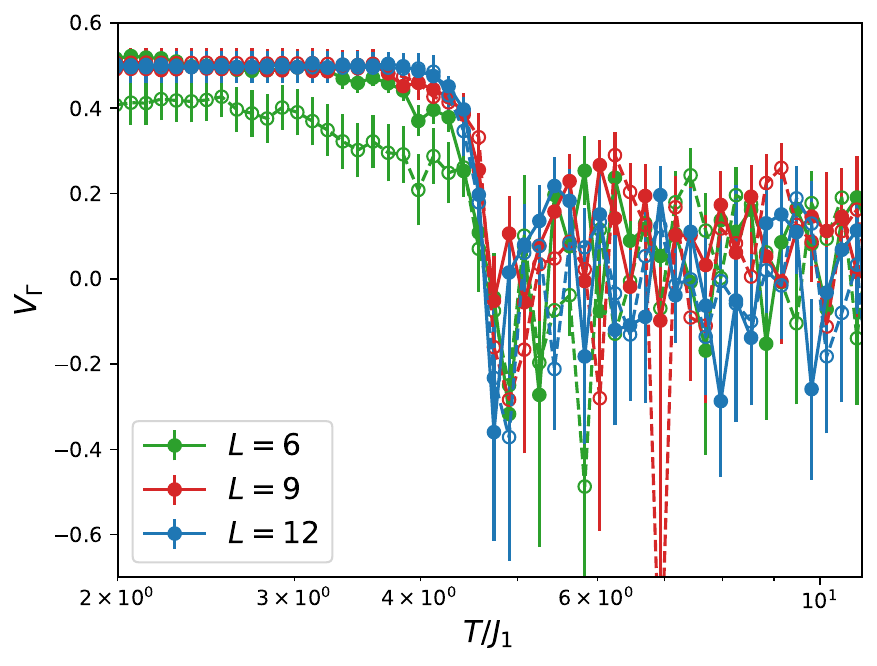}
    \caption{Binder cumulants for the real (filled circles and solid lines) and imaginary (unfilled circles and dashed lines) parts of the order parameter $\Gamma$. The results are shown, from top to bottom, for $J_3'/J_1=0.2$, $J_3'/J_1=0.8$, and $J_3'/J_1=3.2$, for temperatures around the phase transition. The Binder cumulants were computed from $100$ independent simulations, for each of the three system sizes indicated in the legends.}
    \label{fig:Binder}
\end{figure}
%%%
%
%
Examples of $V_\Gamma$ computed for system sizes $L=6,\ 9,\ 12$ are shown in Fig.~\ref{fig:Binder}. In the low-temperature nematic phase $V_\Gamma$ is constant at $1/2$, whereas it fluctuates around zero for temperatures above $T_c$. At or just above the transition there are indications that $V_\Gamma$ becomes negative, similar to the behaviour previously observed for other first order phase transitions in Ising systems~\cite{tsai1998}. There are no indications that the $V_\Gamma$ curves computed for different system sizes cross at a well defined point, as would be the expectation if the transition were second order.

We already noted that the two-peak structure in the energy histograms is most pronounced for $J_3'/J_1=0.8$ (out of the values we considered). We also note that the jump in the order parameter and the Binder cumulant $V_\Gamma$ is sharper at $J_3'/J_1=0.8$ than at other values (see Fig.~\ref{fig:orderpar} and~\ref{fig:Binder}), and the peak in the specific heat capacity is more pronounced (Fig.~\ref{fig:SpecHeat}). One can speculate that the phase transition is sharper when monopoles and triangular lattice excitations have a similar energy, which occurs indeed when $J_3'/J_1 \simeq 0.5$. A more in-depth understanding of this behaviour is left for future studies.

The extent to which we can perform a finite-size scaling analysis of the phase transition is severely limited by the fact that the only system sizes which are numerically accessible to us are $L=3$, $6$, $9$, and $12$. Within these systems sizes one distinct possibility is of course that the first-order two-peak structure of the probability distribution function of the system energy is always present, but for $J_3'/J_1>1$ the peaks are too broad to resolve it. Another, more exotic, possibility could be the presence of multicritical behaviour, as in Ref.~\onlinecite{Jaubert2010}.
%
%
%---------------------------------------------------------

\subsection{Nematic phase and entropy
\label{sec:lowTentropy}} 
As discussed in Sec.~\ref{sec:NSIstates} and in App.~\ref{app:GSbuilding}, we were able to explicitly find a subextensive set of ground states of $\cal{H}$ in Eq.~\eqref{eq:Hamiltonian}, and moreover we provided an argument to construct an extensive number of further ground states from those. Therefore, the Hamiltonian alone is unable to select a zero entropy state at the lowest energy. This argument however does not rule out other selection methods, such as the presence of order by disorder. 

To better understand the nature of the low temperature nematic phase, we numerically computed the integral of the specific heat divided by temperature, to obtain the change in entropy of the system from the known high temperature limit, $k_B \ln(2)$. (We also looked at the behaviour of various local correlators, which however proved somewhat less insightful and are relegated for convenience to App.~\ref{app:correlators}.) 

\begin{figure}
    \centering
    \includegraphics[width=\columnwidth]{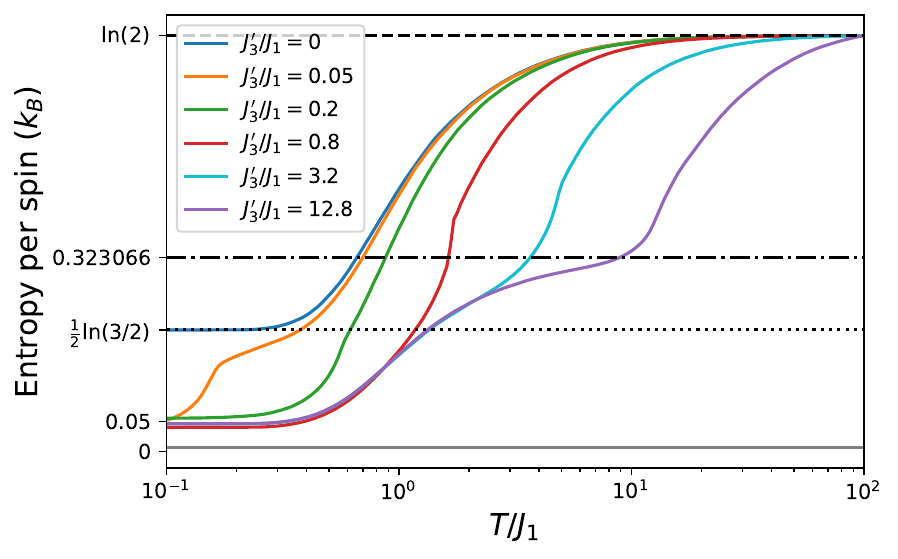}
    \caption{The entropy computed by integrating the specific heat for a system of size $L=9$. The patterned black lines indicate entropy estimates of different models: the infinite temperature entropy per spin, $k_B \ln(2)$, is shown as a dashed line; the dash-dotted line indicates Wannier's estimate of the entropy of the triangular lattice Ising antiferromagnet~\cite{wannier1950}, $0.323066$; and the dotted line indicates the Pauling entropy of conventional spin ice~\cite{pauling1935structure}, $k_B \ln(3/2)/2$. 
    For the system size shown here, $L=9$, we observe a finite residual entropy at the lowest temperature of about $k_B/20$ per spin, for any $J_3'/J_1>0$. This value is significantly higher than the (subextensive) entropy of the fully aligned ground states of $\cal{H}$, which tends to zero as $L^{-1}$ and is indicated by the grey line in the figure. These results are in agreement with a phase transition into an extensive subset of the ice states at low temperature.}
    \label{fig:Entropy}
\end{figure}

The temperature dependence of the entropy of the system is shown in Fig.~\ref{fig:Entropy}, for a range of values of $J_3'/J_1$. 
%
%
%%%
%%%
%
%
Different regimes show the expected plateaux, corresponding to the known spin ice entropy as well as the non-zero entropy of the triangular Ising antiferromagnet. More importantly, we find that the entropy does not drop to zero in the nematic phase, but rather exhibits a robust plateau at the approximate value $k_B/20$. Within the extent of our numerics, this provides evidence that the low temperature phase of our system is indeed extensively degenerate. 
%
%
%----------------------------------------------------------

\subsection{Monopole and triangular defect density
\label{sec:monodens}}
\begin{figure}[ht!]
    \centering
    \includegraphics[width=\columnwidth]{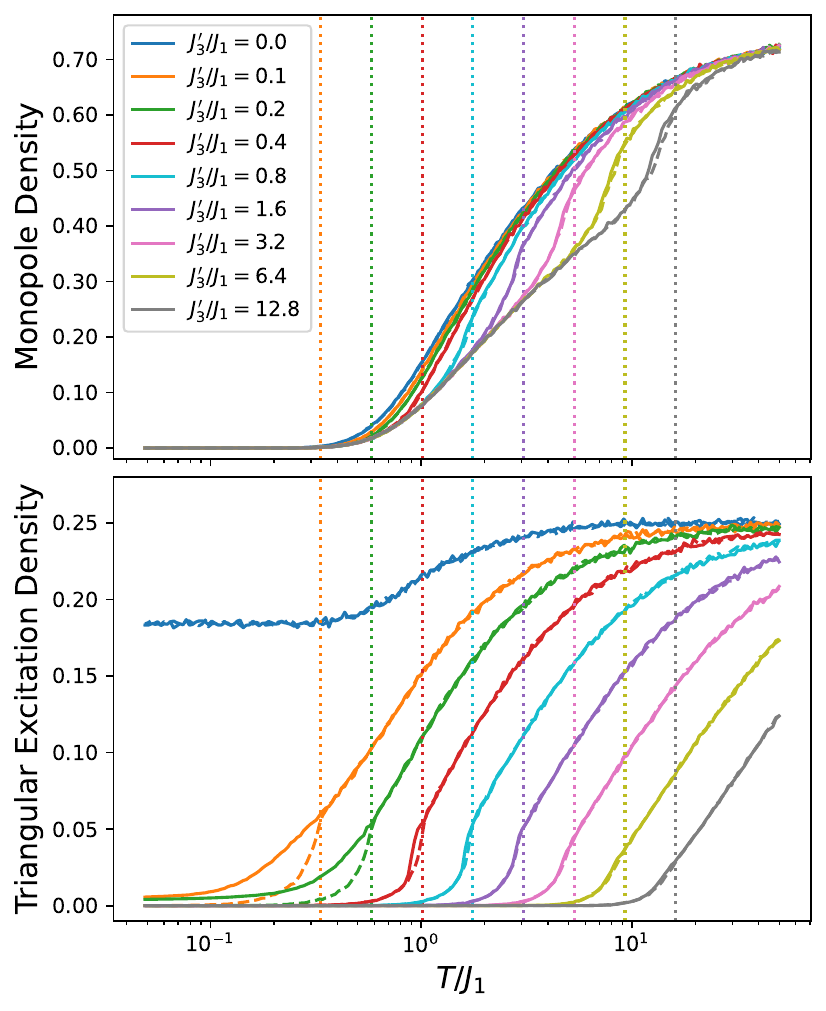}
    \caption{The number density of magnetic monopoles (top) and of triangular lattice excitations (bottom) vs temperature for a range of $J_3'/J_1$ values in a system of size $L=9$. Solid lines indicate simulations where the system was cooled from high temperature, whereas dashed lines indicate simulations begun from a randomly chosen ground state and then heated up. Vertical dotted lines of the corresponding colour indicate the values of $T_c$ obtained from the order parameter (equiv., from the specific heat data).}
    \label{fig:Densities}
\end{figure}

Fig.~\ref{fig:Densities} shows the density of magnetic monopoles and $J_3'$ excitations. 
%
%
%%%
%%%
%
%
Both densities drop when the system enters the nematic phase. However, whereas the monopole density tends to zero at low temperature for all values of $J_3'$, the triangular excitation density does not approach zero for small $J_3'$. There is a simple explanation for this: in a state with zero triangular excitations, monopoles can move along channels of free spins~\footnote{Indeed, the equilibration of the system at low temperatures relies on the code finding such moves (see App.~\ref{app:Methods}).}; however, no corresponding channels exist for the triangular excitations in a zero-monopole state. In a sense, the annihilation of triangular excitations requires the involvement of monopoles -- explaining why the system becomes easily supercooled (akin to a fragile glass~\cite{angell1995}) if the monopole density goes to zero faster than the triangular excitation density (see App.~\ref{app:Supercooling} for additional analysis). 
%
%
%---------------------------------------------------------

\subsection{Summary of the phase diagram
\label{sec:phdiag_summary}}
Considering the behaviour of $\langle \Gamma \rangle$, the specific heat, and the excitation densities, we arrive at the full phase diagram of the model in Fig.~\ref{fig:PhaseDiagram}. Because it is not separated from the paramagnetic phase by any phase transition, it is not strictly correct to speak of a \quotes{spin ice phase}. What we normally refer to when speaking about spin ice is the regime where the ice rules are satisfied on a majority of the tetrahedra, and the crossover from the high-temperature paramagnet to this low-temperature regime is indicated by a colour gradient.

Within the extent of our numerics it appears that a first order transition is present for all values of $J_3'/J_1$ and softens potentially to a critical end point when this ratio diverges. However, we cannot rule out the possibility of a change in nature of the transition, from first order to continuous, at some finite $J_3'/J_1$, or other more exotic scenarios. This shall remain an open question for future studies. 
%
%
%%%%%%%%%%%%%%%%%%%%%%%%%%%%%%%%%%%%%%%%%%%%%%%%%%%%%%%%%%

\section{Anomalous magnetic noise and monopole diffusion
\label{sec:PSD}}
The magnetic noise is computed by measuring the magnetisation of the entire system along some axis and using Welch's method~\cite{welch1967} to compute the power spectral density (PSD), defined as
%%%
\begin{equation}
    \mathrm{PSD}(\omega) = \frac{1}{N_s}\langle |\Tilde{M}(\omega)|^2 \rangle
    \, ,
\end{equation}
%%%
where $\Tilde{M}(\omega)$ is the Fourier transform of the temporal trace of the magnetisation, and $N_s$ is the number of spins in the system.
The system dynamics is simulated by Metropolis type Monte Carlo simulations with single spin-flip updates and a single, temperature independent attempt rate, following the standard model of spin ice dynamics~\cite{jaubert2009,Jaubert2011}.

An exponent $\alpha$ is extracted by fitting the PSD to the function 
%%%
\begin{equation}
    f= \frac{A \, \tau}{1 + (\omega \tau)^\alpha}
    \, .
    \label{eq:fiteq}
\end{equation}
%%%
The noise is said to be anomalous if $\alpha < 2$, where $\alpha=2$ corresponds to the Lorentzian noise of a paramagnet. In experiments on the spin ice compound Dy$_2$Ti$_2$O$_7$, an anomalous exponent $\alpha\approx 1.5$ was observed in the low temperature spin ice regime~\cite{dusad2019, samarakoon2022}. We have since proposed that the unusual dynamical properties of Dy$_2$Ti$_2$O$_7$ are best explained by the existence of dynamical fractals on which monopoles are constrained to move \cite{hallen2022}. These fractal structures arise from the specific combination of dynamical and energetic rules that govern monopole motion. 
However, in Ref.~\onlinecite{samarakoon2022} we also established that a Hamiltonian with a relatively large antiferromagnetic third-neighbour interaction, approximately equivalent to ${\cal H}$ in Eq.~\eqref{eq:Hamiltonian} with $J_3' \gtrsim 0.2 J_1$,  generates anomalous magnetic noise and relaxation times consistent with the experimental observations. Here we connect this observation to the presence of the nematic spin ice phase.

We focus on temperatures where there is a small but non-zero monopole density in the finite size systems considered in our work, namely the regime where the PSD measured in simulations is most clearly anomalous. This constrains us in the choice of temperatures we look at. For simplicity, we take $T/J_1=0.4$ and tune between the phases by considering different values of $J_3'$. At this temperature the phase transition occurs at $J_3'/J_1 \approx 0.13$.

%
%
%%%
\begin{figure}
    \centering
    \includegraphics[width=\columnwidth]{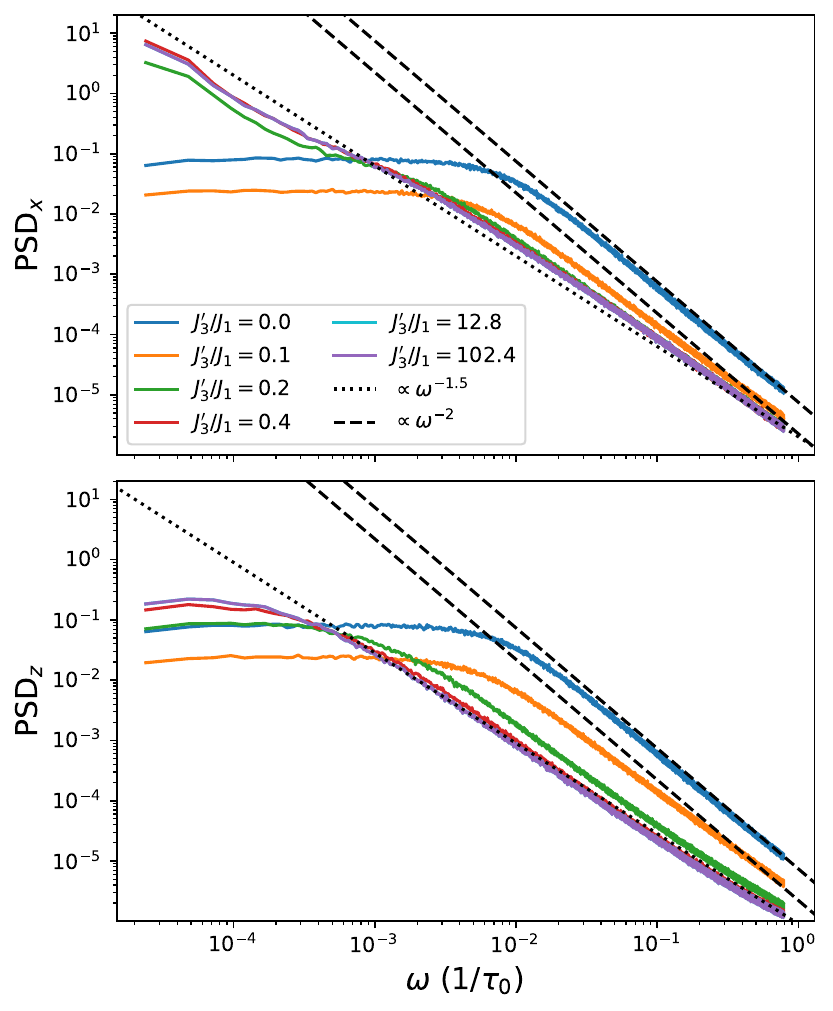}
    \caption{PSD measured along the $x$-direction (top) and the $z$-direction (bottom) at the same temperature $T/J_1 = 0.4$, spanning different values of $J_3'$. The system is in the nematic spin ice phase for $J_3'/J_1 > 0.13$. Note that for $J_3'/J_1 \geq 0.4$ there appears to be no further change in the power law behaviour of the noise with variations in $J_3'$. 
    That the noise is already anomalous for $J_3'/J_1=0.1$ suggests that the anomalous behaviour appears already above the phase transition when $J_3'/T \sim 1$ and the $J_3'$ interactions begin to influence monopole motion. The noise continuously becomes more anomalous across the phase transition; once nematic correlations fully onset, the anomalous behaviour remains unchanged irrespective of how deep one takes the system into the nematic phase by tuning $J_3'$.}
    \label{fig:PSD_varyJ3p}
\end{figure}
%%%
%
%

We observe noise with an anomalous exponent in and near the nematic phase (see Fig.~\ref{fig:PSD_varyJ3p}). Notice that there is no sharp feature in the thermodynamic properties of the system when the noise switches from Lorentzian ($\alpha=2$; well understood for spin ice \cite{samarakoon2022,ryzhkin2005}) to anomalous. The exponent $\alpha$ drops notably below $2$ already in the spin ice phase as the phase transition is approached from above, and it then changes continuously across the phase transition, continuing to decrease inside the nematic spin ice phase before hitting a minimum at $\alpha \approx 1.5$ deep inside the phase. 

In the nematic phase, monopole transport is reduced in one of the three directions, as evidenced by the low frequency plateau present in the lower panel of Fig.~\ref{fig:PSD_varyJ3p}. (Without loss of generality, we take the direction of reduced transport to be the $z$ direction -- consistent with choosing a coordinate system for which the spontaneously selected chain pair is $\beta_{12}$-$\beta_{34}$.)  This is because the paths of free spins are more likely to align with the [110] or [1-10] directions corresponding to the $\beta_{12}$ and $\beta_{34}$ chains. This only occurs below the phase transition. 

Deep in the nematic phase the PSD no longer depends on the value of $J_3'$ (although it still changes with temperature, as this governs the monopole density). It is therefore reasonable to conclude that the anomalous scaling in this regime is caused by the constraints posed on the magnetic monopole motion by the $J_3'$ interactions and by their resulting spin correlations. That the noise does not change when $J_3'$ is increased (above some ratio $J_3'/T \approx 1$) indicates that the monopoles predominantly move along paths of free spins. In the next section we analyse these free paths and the clusters they form. 

From measurements of the spin and $\gamma$ correlations (see Fig.~\ref{fig:spinspincorr} and App.~\ref{app:correlators}), there are no indications of thermodynamic correlations appearing above the phase transition. In the regime above but near the phase transition the energy barriers imposed on the monopoles by the third-neighbour interactions are of similar order as the temperature. These  thus still influence the monopole motion, although no long range correlations have formed. The anomalous exponent observed above the phase transition is most likely a sign of these energy barriers biasing the monopoles to move along `energetically flat' paths. 
%
%
%%%%%%%%%%%%%%%%%%%%%%%%%%%%%%%%%%%%%%%%%%%%%%%%%%%%%%%%%%%

\section{Free paths and clusters
\label{sec:Clusters}}
\begin{figure}
    \centering
    \includegraphics[width=\columnwidth]{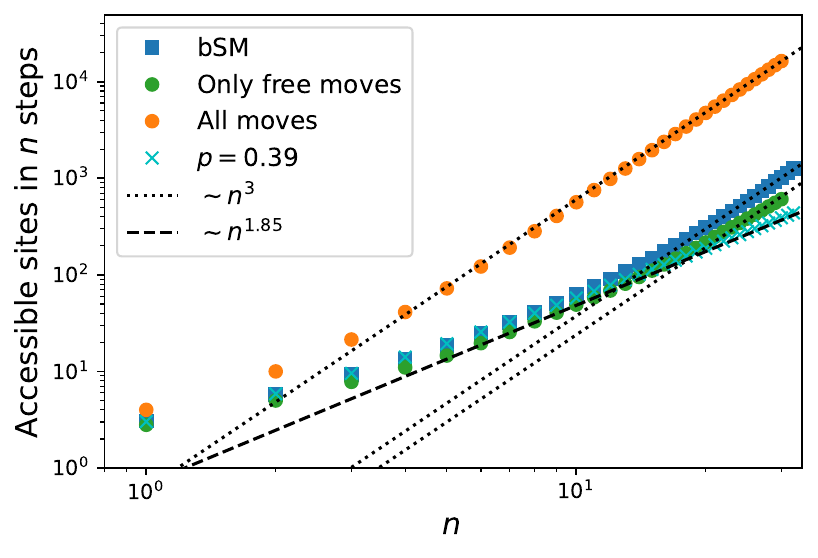}
    \includegraphics[width=\columnwidth]{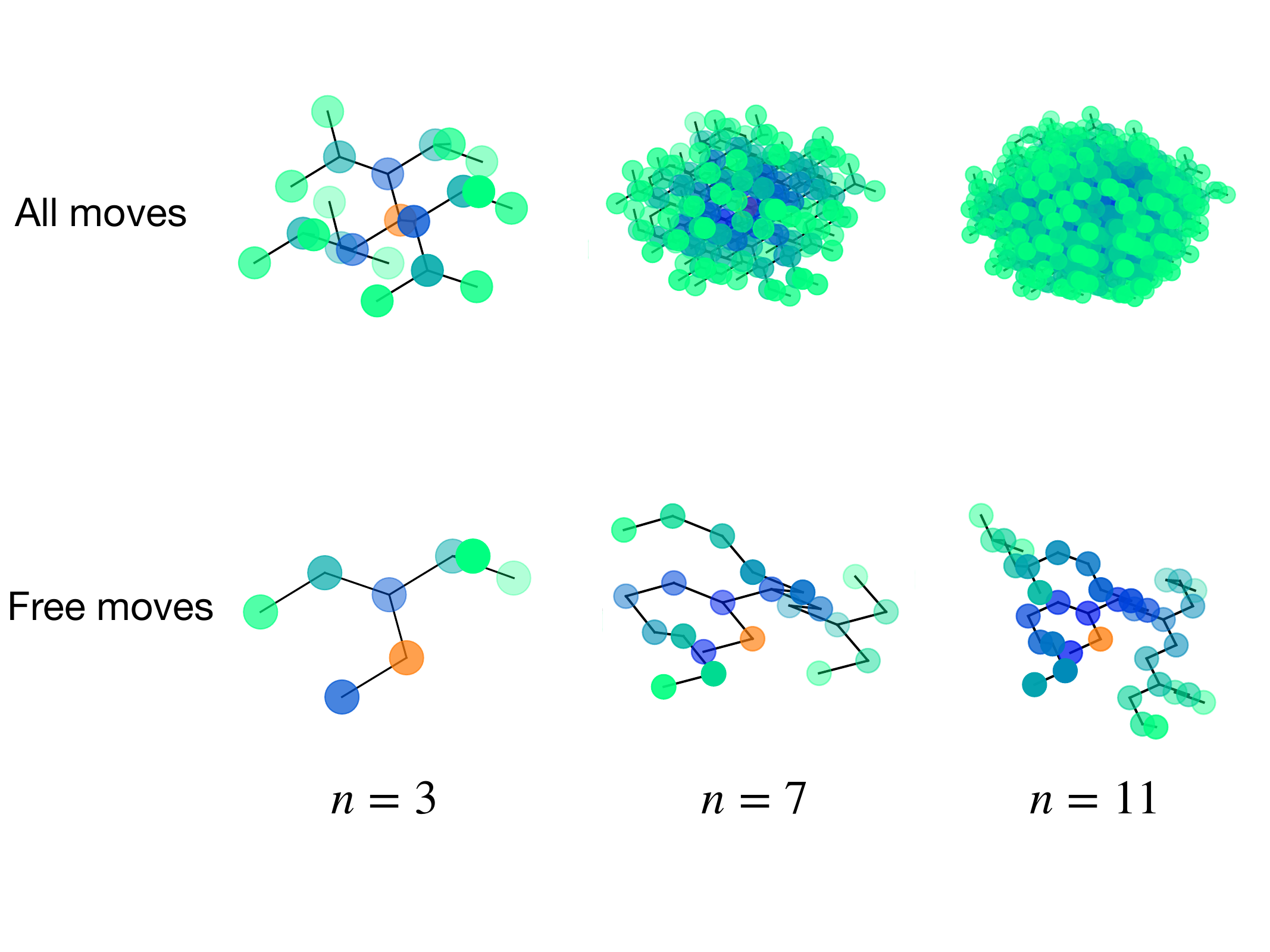}
    \caption{Top: average number of accessible sites for a monopole within chemical distance $n$, computed from 2000 independent configurations. The round markers show the behaviour in the nematic phase when the monopole either makes all moves allowed by the ice rules (orange) or only moves that do not increase the energy (green). The clusters for bSM spin ice~\cite{hallen2022} (no $J_3'$ interaction) and the critical percolation cluster on the diamond lattice are shown for comparison (dark blue squares and light blue crosses, respectively), see main text for details. The agreement between green round markers and light blue crosses suggests that monopoles appear to move on a fractal percolation cluster up to $n \sim 20$. 
    Bottom: Examples of monopole clusters in nematic spin ice when the monopole is either moving without accounting for the $J_3'$ interactions (top row) or when it is only allowed to make moves that do not increase the energy (bottom row).}
    \label{fig:Clusters}
\end{figure}

Following the methods developed in Ref.~\onlinecite{hallen2022} we can map out which sites a monopole can visit within a certain number of steps, see Fig.~\ref{fig:Clusters}. The minimum number of steps a monopole needs to take to reach a site defines the chemical distance $n$ to that site. For every monopole one can then define an \quotes{accessible cluster} of sites that the monopole can reach within chemical distance $n$. We classify a site as accessible if the monopole can move there without at any point creating a double monopole (i.e., a 4-in or 4-out tetrahedron). 

We assume that all flips are attempted with the same rate without dynamical constraints. However, the presence of $J_3'$ interactions has two effects: (i) it introduces energy barriers associated with the monopole moves; and (ii) it alters the spin correlations, thereby altering the spatial distribution of spins across which a monopole cannot hop (so-called `blocked spins' in SM dynamics~\cite{hallen2022}, which incur an energy cost of order $J_1$). There remain a significant number of spins that can flip without any $J_3'$ energy cost, and in the nematic phase these form clusters on which the monopoles predominantly move (for $J_3'\gtrsim T$). Examples of such clusters are shown in Fig.~\ref{fig:Clusters}.
%
%
%%%
%%%
%
%

By averaging over a large number of different nematic spin ice configurations with monopoles, we can look at how these clusters grow with increasing $n$. The size of the clusters formed by moving only through free spins appears to grow with a fractal exponent. This is likely the explanation for the observed anomalous magnetisation noise. Intriguingly, for $n \lesssim 20$ the scaling of the cluster size, shown in Fig.~\ref{fig:Clusters}, closely follows that of the percolating cluster at the critical filling fraction in the bond percolation model on the diamond lattice (i.e., at concentration $p \approx 0.39$ of bonds, as noted in Fig.~\ref{fig:Clusters}). For length scales greater than $n\sim 20$ the system crosses over to the conventional three-dimensional behaviour of a regular lattice, indicating that the energetic constraints on monopole motion effectively place the clusters they move on slightly above the critical percolation threshold~\cite{stauffer2018}. It is the distance to the critical percolation threshold that fixes the length and time scale on which monopoles exhibit anomalous motion. These clusters in nematic spin ice behave similarly to the fractal clusters formed by the combination of dynamical and energetic constraints in the so-called \quotes{beyond the standard model} (bSM) dynamics of spin ice~\cite{hallen2022}. The clusters formed by free spins in the nematic spin ice model are however not isotropic; they grow more slowly with increasing $n$ in the direction normal to the selected chain pairs in the given nematic order. 

\begin{figure}
    \centering
    \includegraphics[width=\columnwidth]{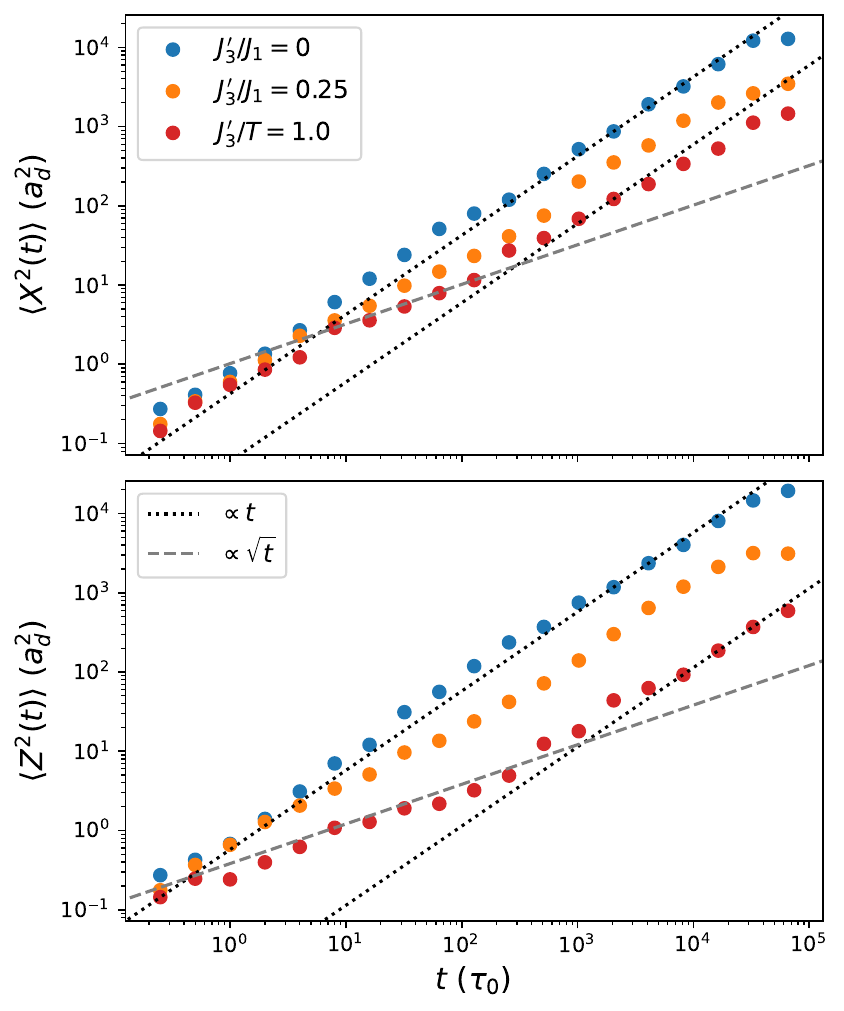}
    \caption{Mean-squared displacement of individual monopoles for a range of values of $J_3'$. The incipient plateau appearing at long times is a finite-size effect (this is a system with $L=15$). $\langle Y^2(t) \rangle$ behaves like $\langle X^2(t) \rangle$, whereas $\langle Z^2(t)\rangle$ clearly grows more slowly, as one would expect from the behaviour of the PSD. In these simulations a single monopole pair is created in each realisation, and only one of the two monopoles is allowed to move (see App.~\ref{app:Methods}).}
    \label{fig:MSD}
\end{figure}

The non-isotropic clusters explain why the noise is reduced along the $z$-axis: there are generally fewer opportunities for monopoles to travel in the $z$-direction and the extension of the clusters is correspondingly smaller, creating a low frequency plateau which does not appear in the $x$- and $y$-components (see Fig.~\ref{fig:PSD_varyJ3p}). The fractal nature of the clusters explains why the noise is anomalous: Particles performing a random walk on a fractal graph spread sub-diffusively, with a mean-squared displacement $\langle R^2(t) \rangle \sim t^{\sigma}$ and $\sigma<1$~\cite{deGennes1976,Havlin1987,hughes2021}. The fluctuations of the system magnetisation is directly proportional to the random displacement of the monopoles, and the PSD of a random walker on a fractal cluster is ${\rm PSD} \sim \omega^{-(1+\sigma)}$~\cite{hallen2022}. That the monopole motion is indeed sub-diffusive, on times $t \lesssim 10^3 \tau_0$, can be seen from measurements of the mean-squared displacement for individual monopoles (see Fig.~\ref{fig:MSD}). 
%
%
%%%
%%%
%
%
On longer time scales we observe a crossover to normal diffusion, consistent with a system that only exhibits fractal scale invariance on sufficiently short length scales.
%
%
%%%%%%%%%%%%%%%%%%%%%%%%%%%%%%%%%%%%%%%%%%%%%%%%%%%%%%%%%%%

\section{Conclusion
\label{sec:conclusions}}
That the introduction of specific farther-neighbour interactions, on top of the nearest-neighbour spin ice Hamiltonian, does not lead to long range magnetic order at zero temperature, but instead selects an \emph{extensive subset} of the spin ice states, was a priori highly unexpected. In this respect, the nematic spin ice phase discussed here is crucially different from the nematic states argued in quantum spin ice models~\cite{taillefumier2017}.

Although the choice of the two exchange interaction paths is fairly specific, the relative strength of the two types of interactions does not need to be fine-tuned. The nematic spin ice phase appears below some $T_c$ for any choice of $J_1 > 0$ and $J_3' > 0$. The exact characterisation and counting of the extensively many ground states of ${\cal H}$ remains a possible avenue for further work. And so is the fate of the first order transition in the limit $J_3' \gg J_1$. 

The most detailed characterisation of the Dy$_2$Ti$_2$O$_7$ spin Hamiltonian predicts that the third-neighbour exchange interaction across the hexagons is indeed the dominant farther-neighbour exchange term, with an effective $J_3'/J_1$ on the order of $0.1$~\cite{samarakoon2020}. However, Dy$_2$Ti$_2$O$_7$ freezes below approximately $0.55$~K~\cite{snyder2004} and temperatures where any nematic spin ice physics may manifest itself are not experimentally accessible. It is not out of the question, however, that other spin ice materials could be devised, e.g., by chemical substitution, that approximately realise ${\cal H}$ and enter a nematic spin ice phase. An alternative avenue for realising this phase are artificial spin ice systems, where the geometry and interaction strengths may be more easily tailored to realise our model.

We note in passing that the study of pyrochlore magnets with further range interactions has a long history, reaching at least as far back as the mean-field analysis of Reimers, Berlinsky and Shi~\cite{reimers1991mean}. Many interesting phenomena have been unearthed along the way, such as the appearance of correlations mimicking  hexagonal cluster arrangements~\cite{yavors2008,conlon2010absent}, or very recently, the appearance of sublattice-paired spirals~\cite{Glittum2023}. Further studies of such models continue to appear promising.

More broadly, the discovery of the nematic spin ice model -- driven, as it was, by the search for anomalous dynamical behaviour -- provides yet another example of how the \emph{frequency dependence} of fluctuations can contain information about the \emph{spatial correlations} in a system. In nematic spin ice the magnetic monopole quasiparticles are sub-diffusive due to correlated energetic constraints on their motion, effectively biasing the monopoles to move on clusters with fractal properties. Although their origin is different, these \quotes{energetic fractals} are similar to the dynamical fractals believed to cause anomalous magnetic noise in Dy$_2$Ti$_2$O$_7$~\cite{hallen2022} -- explaining why simulations with ${\cal H}$ in Eq.~\eqref{eq:Hamiltonian} closely resemble the experimentally observed magnetic noise. The combination of the two effects is a promising direction for further work, as is the further characterisation of the anisotropic free clusters and their potential connection to anisotropic percolation models~\cite{redner1979}.
\\

\noindent All numerical data presented in this paper are available online \cite{*[{All data presented in this paper are available at: }][] Data}.
%
%
%%%%%%%%%%%%%%%%%%%%%%%%%%%%%%%%%%%%%%%%%%%%%%%%%%%%%%%%%%%

\section*{Acknowledgements}
We are grateful to Peter Holdsworth and Ludovic Jaubert for useful discussions; and to Santiago Grigera and Alan Tennant for collaboration on related work.
For the purpose of open access, the authors have applied a Creative Commons Attribution (CC BY) licence to any Author Accepted Manuscript version arising from this submission. 
This work was supported in part by the Deutsche Forschungsgemeinschaft under grant SFB 1143 (Project-ID No. 247310070); by the cluster of excellence 
ct.qmat (EXC 2147, Project-ID No. 390858490); and by the Engineering and Physical Sciences Research Council (EPSRC) grants No. EP/P034616/1, EP/V062654/1 and EP/T028580/1. 

%%%%%%%%%%%%%%%%%%%%%%%%%%%%%%%%%%%%%%%%%%%%%%%%%%%%%%%%%%%
\appendix
%%%%%%%%%%%%%%%%%%%%%%%%%%%%%%%%%%%%%%%%%%%%%%%%%%%%%%%%%%%

\section{Pauling estimate of the ground state entropy
\label{sec:pauling}}
The ground state entropy of conventional spin ice can be estimated using simple arguments developed by Linus Pauling for H$_2$O ice~\cite{pauling1935structure}. The argument runs as follows: Every tetrahedron has $2^4=16$ possible spin configurations of which 6 are 2-in 2-out configurations that minimise the energy locally. In a spin ice system with $N_s$ spins there are $2^{N_s}$ spin configurations. Assuming that the constraints from the $N_t=N_s/2$ tetrahedra can be treated independently, the number of ground states can be estimated as 
%%%
\begin{equation}
    \Omega = 2^{N_s} \left(\frac{6}{16} \right)^{N_s/2} = \left(\frac{3}{2} \right)^{N_s/2}
    \, .
\end{equation}
%%%
The corresponding ground state entropy is ${\cal S} = k_B \ln{\Omega} = \frac{k_B N_s}{2} \ln{\frac{3}{2}}$, and it is a lower bound for the entropy of the ensemble of 2-in 2-out states. 

Extending the Pauling estimate to our model, Eq.~\eqref{eq:Hamiltonian}, we similarly treat all of the ``up'' triangles formed by the third-neighbour interactions as independent constraints on the ground state. There are $N_s$ such triangles, each with $2^3=8$ spin configurations of which $6$ minimise the energy on the triangle. The total number of ground states is then estimated as 
%%%
\begin{equation}
    \Omega = 2^{N_s} \left(\frac{6}{16} \right)^{N_s/2} \left( \frac{6}{8} \right)^{N_s} = \left( \frac{27}{32}\right)^{N_s/2}
    \, ,
\end{equation}
%%%
resulting in a corresponding entropy 
%%%
\begin{equation}
    {\cal S} = \frac{k_B N_s}{2} \ln{\frac{27}{32}} \approx -0.08 k_B N_s
    \, . 
\end{equation}
%%%
This na\"\i ve estimate thus fails to capture the extensive degeneracy of the nematic spin ice ground state, as it predicts a negative entropy consistent with a conventionally ordered ground state. 
%
%
%%%%%%%%%%%%%%%%%%%%%%%%%%%%%%%%%%%%%%%%%%%%%%%%%%%%%%%%%%%

\section{Commensurability and system size\label{app:SystemSize}}
The construction of nematic ground states described in the main text and in App.~\ref{app:GSbuilding} relies on the triangular planes wrapping around the system properly (with periodic boundary conditions), so that the sublattices of the triangular lattice are well defined. This is true when our system size $L$ (i.e., a cubic system formed by $L^3$ cubic unit cells, each containing $16$ spins) is an integer multiple of three. The triangular planes are of size $2L\times 2L$, and the sublattice structure of the tripartite triangular lattice can thus be respected by the periodic boundary conditions.

In this work we have chosen to only work with system sizes compatible with the triangular sublattices. We have found strong finite-size effects when using other system sizes.

%%%%%%%%%%%%%%%%%%%%%%%%%%%%%%%%%%%%%%%%%%%%%%%%%%%%%%%%%%%
%%%%%%%%%%%%%%%%%%%%%%%%%%%%%%%%%%%%%%%%%%%%%%%%%%%%%%%%%%%

\section{Numerical methods
\label{app:Methods}}
All numerical data presented in this paper were obtained through Monte Carlo simulations using the Metropolis algorithm. We have used periodic boundary conditions in all cases. The primary method used is simulated annealing, where the system is initialised in a random configuration at high temperature and then cooled gradually. For comparison (or in some cases in order to avoid supercooling -- see below) simulations were initiated in randomly selected ground states and the temperature was then gradually increased instead. In App.~\ref{app:GSbuilding} we explain how groundstates can be generated directly.

In conjunction with simple single-spin flip updates, we have made use of several different cluster updates to speed up equilibration. Any closed loops formed by spins aligned head-to-tail can be flipped without the creation or annihilation of magnetic monopoles (i.e., without incurring any $J_1$ energy cost). Flipping such loops therefore does not change the system energy if $J_3'=0$, and is a standard method for simulating spin ice systems at low temperatures~\cite{barkema1998, melko2001}. The loops are found by randomly selecting a starting spin and direction, and then forming a path by randomly selecting a neighbouring spin with the correct sign at every step. When this path encounters a previously visited site a closed loop has been identified (and the residual dangling path is discarded). Alternatively, one can allow the walk to continue until the path reaches the starting site, which forms a loop containing (generally) more spins but which also takes longer time. Both types of loop updates were used here. 

One can move monopoles over large distances by identifying a path of aligned spins which terminates when the path encounters a monopole (of the correct sign). The paths are generated in the same way as for the closed loop updates. Flipping all the spins along the path effectively moves the monopole to the starting site of the path, and this happens at zero energy cost if $J_3'=0$. Both these monopole moves and the closed loop updates are most useful to equilibrate the system in the regime where the monopole density is relatively low.

If $J_3'>0$, which is the case we are primarily concerned with here, the cluster updates described above will generally change the system energy by creating and annihilating triangular lattice excitations. The standard spin ice updates therefore become progressively less useful when $T$ is lowered below $J_3'$, as flipping a path or loop selected without any consideration for the $J_3'$ interactions is likely to carry a large energy penalty. However, a significant proportion of the spins can flip at zero $J_3'$ energy cost (as explained in the main text and in App.~\ref{app:GSbuilding}) and we also make use of closed loop and monopole move updates which only include such free spins. Flipping a loop or path of aligned and $J_3'$-free spins never changes the system energy, and these can be found by simply limiting the path-finding random walks to only include $J_3'$-free spins.

Finally, we used a further cluster update that focuses solely on the triangular planes and $J_3'$ interactions. For the triangular lattice Ising antiferromagnet, one can find entire blocks of spins that can flip together without changing the energy. This is done by defining a dimer representation of the triangular lattice Ising model, which lives on the dual hexagonal lattice. The links of the latter are bisectors of the bonds between the spins, and dimers are placed where bonds connect spins of the same sign. Three links of the hexagonal lattice meet at the centre of every triangle, and if there is no excitation at the triangle only one of these has a dimer. One can find closed loops of alternating filled and empty links through a similar random walk as those used to find loops of aligned spins in spin ice. If one flips all of the spins inside such a dimer loop, it does not change the number of triangular excitations. Dimer representations can be defined for each of the $4L$ interpenetrating triangular lattice planes of the pyrochlore lattice, and they can then be used to identify sets of spins that can flip without changing the $J_3'$ energy. These updates do not conserve the number of magnetic monopoles, and are therefore most useful when $J_3' > T > J_1$.

Only single spin flip updates were used for the dynamical measurements of the magnetisation noise and monopole mean-squared displacement. (Cluster updates were used to prepare the system in thermodynamic equilibrium at the desired temperature before a dynamical measurement was begun.)  One Monte Carlo sweep consists of $N_s$ attempted spin flips, where $N_s$ is the number of spins in the system; each spin to be flipped is selected at random. One sweep corresponds to the characteristic time scale $\tau_0$~\cite{jaubert2009} ($\tau_0=1$ in typical MC units). 

The dynamical measurements used to compute the mean-squared displacement shown in Fig.~\ref{fig:MSD} were done allowing only a single monopole to move. These simulations are initiated from a ground state of the model, and loop updates which do not create monopoles, but may create triangular excitations, are then performed to scramble the system. The only relevant scale for these updates is $J_3'/T$, and depending on the choice of this ratio the loop updates may erase the initial nematic correlations or not. A single monopole pair is then created by flipping a randomly selected spin, and annihilation of this pair is thereafter forbidden. The Monte Carlo update now consist of randomly selecting one of the four spins surrounding the positive monopole and flipping this with the usual Metropolis probability $\min\left[1, \exp{\left(-\Delta E / T\right)}\right]$ if the spin is a majority spin (i.e., we only allow moves that hop the positive monopole). This is a good approximation of \quotes{natural} monopole dynamics under the SM (the standard model of spin ice dynamics; see main text and Ref.~\onlinecite{jaubert2009} and \onlinecite{Jaubert2011}) for $T\ll J_1$, where the monopole density is low and creation and annihilation events are rare. Four such attempted spin flips correspond to time $\tau_0$. Before any measurement was begun, $10N_s$ updates were performed to ensure that the positive monopole has a chance to separate from its negative partner in the pair-creation event.

We have not been able to devise any cluster updates which efficiently transport triangular excitations without the creation of magnetic monopoles. The numerical equilibration time therefore becomes very long in the regime $T<T_c$ if $J_3'/J_1$ is small. More precisely, this occurs in the regime where the density of monopoles in the system becomes much smaller than the density of triangular excitations, and this is the reason why we observe supercooling for small but nonzero $J_3'/J_1$ (see Fig.~\ref{fig:Densities} and App.~\ref{app:Supercooling}). 
%
%
%%%%%%%%%%%%%%%%%%%%%%%%%%%%%%%%%%%%%%%%%%%%%%%%%%%%%%%%%%%

\section{Constructing ground states
\label{app:GSbuilding}}
The pyrochlore lattice can be divided into four sublattices, with every tetrahedron containing one spin from each sublattice. Spins on sublattices $1$ to $4$ respectively point along their local easy-axis directions, which can be defined as $\Vec{e}_1 = \frac{1}{\sqrt{3}}(-1,-1,-1)$, $\Vec{e}_2 = \frac{1}{\sqrt{3}}(1,1,-1)$, $\Vec{e}_3 = \frac{1}{\sqrt{3}}(1,-1,1)$, and $\Vec{e}_4 = \frac{1}{\sqrt{3}}(-1, 1, 1)$, in the global crystallographic reference frame. 

A subextensive set of long range ordered ground states for the Hamiltonian in Eq.~\eqref{eq:Hamiltonian} can be constructed as follows. 
Consider the three possible ways of pairing the four sublattices; for instance, let us focus without loss of generality on $12$ and $34$. All the sites belonging to sublattices $1$ and $2$ form straight lines across the system, and so do the sites belonging to sublattices $3$ and $4$. Alignment of spins along these chains occurs in conventional spin ice under the influence of a magnetic field in the [110] direction~\cite{fennell2002field}. The chains are conventionally called $\alpha$- or $\beta$-chains, depending on their direction relative to the applied field~\cite{hiroi2003ferromagnetic}.
Here, where no external field is applied, we refer to all chains as $\beta$-chains; the $\beta_{12}$ chains lie along the [110] direction and the $\beta_{34}$ chains along the [-110] direction, and both of them are orthogonal to the $z$ crystallographic axis (i.e., [001]).
In a system of linear size $L$ (which has $16L^3$ spins) there are $2L^2$ chains of each type, and every spin in the system belongs to exactly one of these. These chains are illustrated in Fig.~\ref{fig:Chains}.

In addition to $\beta_{12}$ and $\beta_{34}$ chains, there are two other possible pairings: $\beta_{13}$ along [101] and $\beta_{24}$ along [-101] or $\beta_{14}$ along [011] and $\beta_{23}$ along [0-11]. The three choices of chain pairs are related by simple rotations, and we can focus on the first choice above without loss of generality.

Any state where the spins along each chain are aligned head to tail is a spin ice state (they minimise the nearest-neighbour interaction energy). Since the polarisation of each chain can be chosen independently, there are $3\times 2^{4L^2}$ such ground states (with the factor of 3 accounting for the three possible choices of chain pairs).

\begin{center}
    \begin{figure}
        \centering
        \includegraphics[width=0.9\textwidth]{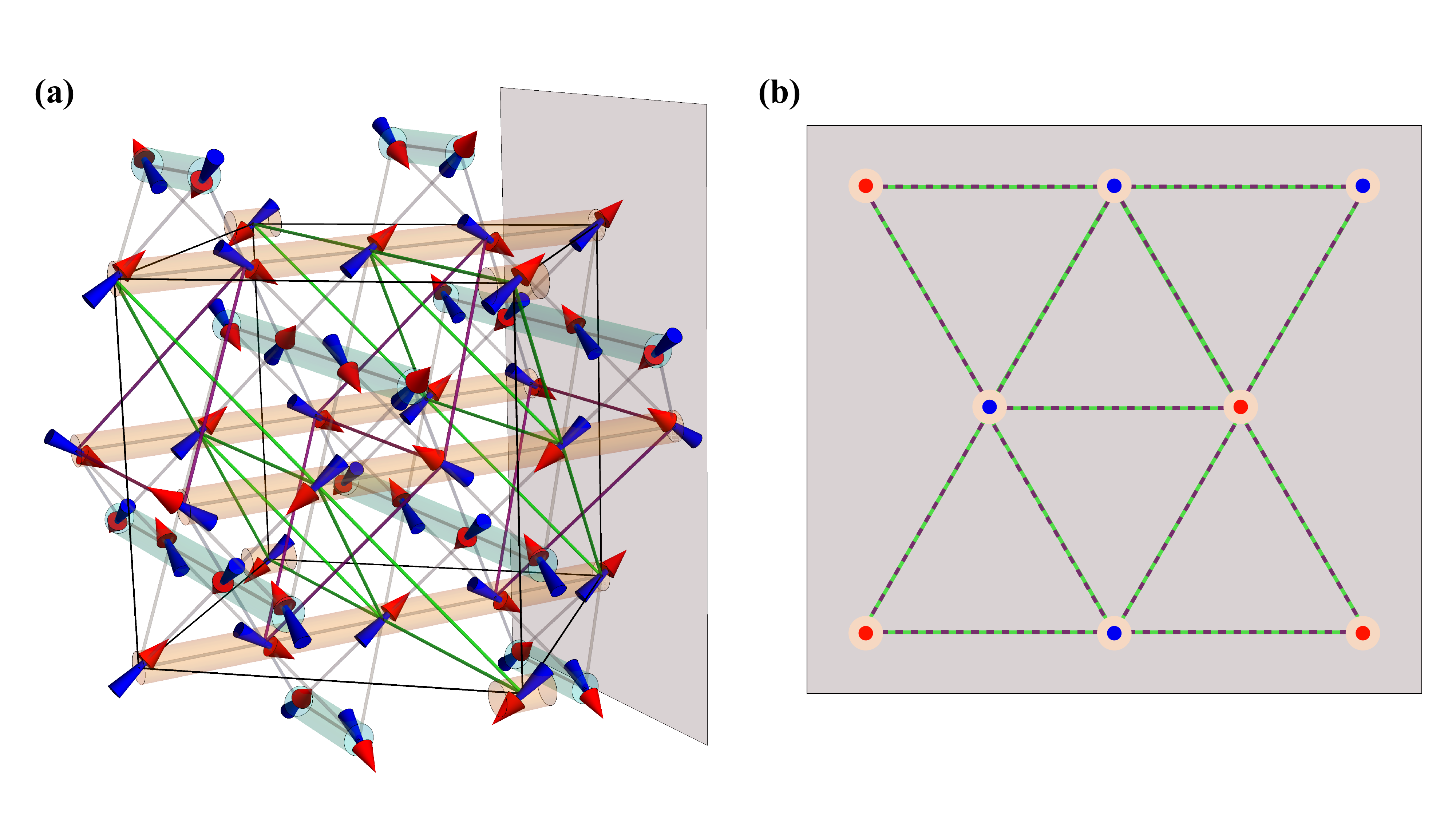}
        \caption{(a) Section of spin ice in a long-range ordered ground state of ${\cal H}$ with spins aligned along the $\beta_{12}$ (orange) and $\beta_{34}$ (cyan) chains. The $J_3'$ interactions are shown for spins on sublattice 1 (green lines) and sublattice 2 (purple lines). If one projects the chains onto the plane normal to the chain direction (shown in grey for the $\beta_{12}$ chains), one is left with a triangular pattern as shown in panel b.
        (b) Combined triangular pattern from the projection of the $\beta_{12}$ chains. A red dot indicates that the spins along this chain point out of the plane and a blue dot indicates that they point into the plane -- one can think of each chain as a net Ising spin. Each bond on the combined pattern represents $4L$ $J_3'$ interactions in spin ice -- including both the ones between spins on sublattice 1 (green lines in (a)) and between spins on sublattice 2 (purple lines in (a)). Directing the chains so that each triangle on the triangular pattern has two blue and one red dots or one blue and two red dots therefore minimises the energy of all the $J_3'$ planes formed by spins on sublattice 1 and 2.}
        \label{fig:Chains+TriangleProjection}
    \end{figure}
\end{center}

If we now look at one set of $\beta$-chains, say $\beta_{12}$, and project them onto the plane perpendicular to the chain direction, we can see that they form a triangular pattern (see Fig.~\ref{fig:Chains+TriangleProjection}). When projected in this way, every $\beta_{12}$ chain can be treated as an effective Ising spin $\eta^{12}$ and the $J_3'$ interactions, which of course connect nearest-neighbour $\beta_{12}$ chains, make these effective Ising spins interact with a nearest-neighbour antiferromagnetic coupling of strength $4LJ_3'$, where $4L$ is the number of spins on each chain. The equivalent projection to a plane can be performed for the $\beta_{34}$ chains, resulting in a completely independent set of effective Ising spins $\eta^{34}$, which also form a triangular pattern with antiferromagnetic nearest-neighbour interactions between the effective spins. All $J_3'$ interactions of ${\cal H}$, Eq.~\eqref{eq:Hamiltonian}, are exactly accounted for by the interactions between the effective Ising spins, and, as polarised chains automatically minimise the $J_1$ energy, the Hamiltonian can now be expressed as 
%%%
\begin{widetext}
\[
    {\cal H} = J_1 \sum_{\langle i, j\rangle} \sigma_i \sigma_j + J_3' \sum_{\langle i, j\rangle_{3'}} \sigma_i \sigma_j =
    - 16 L^3 J_1 
    +
    4 L J_3' \sum_{\langle k, l \rangle_{12}} \eta_k^{(12)} \eta_l^{(12)} 
    +
    4 L J_3' \sum_{\langle k, l \rangle_{34}} \eta_k^{(34)} \eta_l^{(34)} 
    \, ,
\]
\end{widetext}
%%%
where the final two sums are over all nearest-neighbour pairs on the triangular patterns formed by the $\beta_{12}$ and $\beta_{34}$ chains respectively. Any configuration where every triangle on the triangular patterns has two effective spins pointing up and one down or two down and one up will have $J_3'$ energy $2\times 4LJ_3' \times (-2L^2) = -16L^3 J_3'$. This is the minimum possible energy of the $J_3'$ interactions, and any such configuration is thus a ground state of our model. 

%
%
%%%
\onecolumngrid
\twocolumngrid
%%%
%
%

A sub-extensive set of polarised groundstates can be constructed following a simple argument that we outline hereafter.
The standard triangular lattice has three sublattices, that we call $A$, $B$ and $C$. Ground states of the triangular lattice Ising antiferromagnet can, for example, be constructed by making all spins on the $A$ sublattice point up and all spins on the $B$ sublattice point down, and the spins on the $C$ sublattice can then be chosen at random~\cite{wannier1950, Houtappel1950}. 
Similarly, we can divide the set of all $\beta_{12}$ chains, or equivalently the effective Ising spins $\eta^{(12)}$, into the three triangular pattern sublattices. Let all chains on the first sublattice point in the same direction, let all chains on the second sublattice point in the opposite direction, and finally randomly choose the direction of each chain on the third sublattice (all this while keeping all spins in each given chain aligned head to tail). We independently perform the same process for all the $\beta_{34}$ chains as well. Any resulting configuration is guaranteed to be a ground state of the model.

The total number of such ordered ground states is given by 
%%%
\begin{equation}
    3\times \left(6 \times 2^{2L^2 / 3} \right)^2 = 108\times 2^{4L^2/3}
    \, .
\end{equation}
%%%
The first factor of $3$ comes from the three possible choices of chain pairs. The factor of $6$ comes from selecting which of the three triangular sublattices is all up and which is all down. The factor of $2^{2L^2/3}$ comes from the fact that there are $2L^2/3$ chains on the remaining (undetermined) sublattice, and each of these can be independently chosen to point in a random direction. Finally, the overall square accounts for the fact that the $\beta_{12}$ and $\beta_{34}$ chains are independent. This is a lower bound on the number of chain-polarised ground states; in many of these states there will be chains on sublattice $A$ or $B$ that are also free to flip at no energy cost. These \quotes{accidentally flippable} chains increase the number of possible states.

We have described how a subextensive set of ground states can be generated for our Hamiltonian. These have long range order. However, as we argue in the main text, there is in fact an extensive number of exact ground states for this model. To understand how we can generate further ground states from the perfectly ordered chain states described above, we return to the triangular lattice Ising antiferromagnet. The spins on the $C$ sublattice are free to choose either of their two states without any changes to the energy. Similarly, any of the free chains in our ordered ground states can reverse direction at no energy cost. However, it is not necessary to flip the entire chain. Flipping a finite segment of the chain does not change the $J_3'$ energy, but does generate a pair of magnetic monopoles at each end of the segment. This can be avoided if one selects segments of freely flipping chains of both the $\beta_{12}$ and the $\beta_{34}$ types such that the segments form a closed loop of spins aligned head to tail. This loop can then be flipped at no energy cost.

The $\beta_{12}$ and $\beta_{34}$ chains are perpendicular to each other and meet at tetrahedra (see Fig.~\ref{fig:Chains}). As one third of the chains of each type are free to flip, there are a large number of crossing points between freely flippable chains in an ordered ground state. We can consider a thin sheet, extending through the system in the $x$ and $y$ directions (i.e., [100] and [010]), and of width of one tetrahedron in the $z$-direction. Such a sheet contains $L$ $\beta_{12}$ chains and $L$ $\beta_{34}$ chains. At least every third $\beta_{12}$ chain and every third $\beta_{34}$ chain in the sheet is free to flip. If the sheet contains at least one free $\beta_{12}$ chain polarised in each direction and at least one free $\beta_{34}$ chain polarised in each direction, it will be possible to form a free loop inside the sheet. In the majority of states formed with $L\geq 6$ there will therefore be a non-zero number of free loops. 

(For completeness, we ought to mention that for $L=3$, each of the sheets described above only contains one free chain of each type and there are no non-winding free loops in the ordered ground states. However, with simulated annealing we have nonetheless found other ground states for $L=3$ that do have non-winding free loops.) 

For any model with only short-ranged interactions, a ground state of a system with side length $L=nl$ (for integer $n$ and $l$) can be formed by tiling a ground state configuration (under periodic boundary conditions) of a smaller system of side length $l$, $n$ times. If the system of side length $l$ has a minimum of two ground states connected by a local update which does not affect the boundaries, these updates can be performed independently in the larger system without changing the system energy. A system of size $L^d$ (in $d$ dimensions) thus has a minimum of $2^{L^d / l^d}$ ground states -- and it is extensively degenerate (the number of states grows exponentially with the system volume). 
The existence of non-winding free loops in the ordered ground states thus proves that our model has an extensive number of ground states.

Finally, it is worth noting that although the free loop flips break the long range order along the chains, they do not restore rotational symmetry. Ground states found through simulated annealing from high temperature do not display long range order along the chains, but do display stronger correlations along one pair of $\beta$-chains and thus break rotational symmetry. (This is true for all $\sim 10^4$ ground states found in this way.) We should note that we are not able to prove that all possible ground states of ${\cal H}$ display rotational symmetry breaking, and there remains a possibility that we are observing a strong order-by-disorder effect.
%%%%%%%%%%%%%%%%%%%%%%%%%%%%%%%%%%%%%%%%%%%%%%%%%%%%%%%%%%%
%%%%%%%%%%%%%%%%%%%%%%%%%%%%%%%%%%%%%%%%%%%%%%%%%%%%%%%%%%%

\section{Spin and $\gamma$ correlators \label{app:correlators}}
Further insight into the nematic spin ice phase and the corresponding phase transition can be obtained from real space correlators like $(-1)^{m}\langle \sigma(0) \sigma(\mathbf{r}) \rangle$ (discussed briefly in the main text) and $\langle \gamma({\bf r}) \gamma({\bf r}') \rangle$. We remind the reader that $m$ is an integer equal to the number of steps between $\mathbf{0}$ and  $\mathbf{r}$ and $\gamma(\mathbf{r})$ is the Potts variable $\gamma_t$ in Eq.~\eqref{eq:PottsVar} at the position of the $t^{\rm th}$ tetrahedron: ${\bf r} = {\bf r}_t$. 

\begin{center}

\begin{figure}[h!]
    \centering
    \includegraphics[width=0.45\columnwidth]{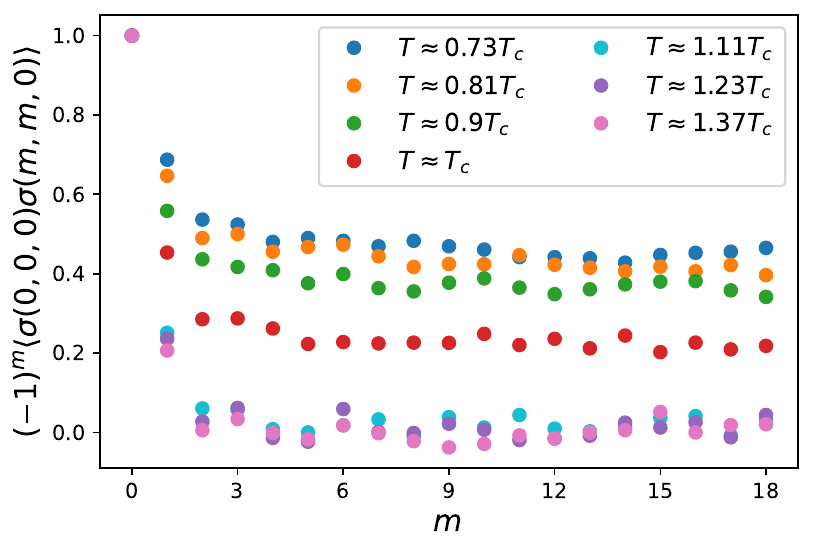}
    \includegraphics[width=0.45\columnwidth]{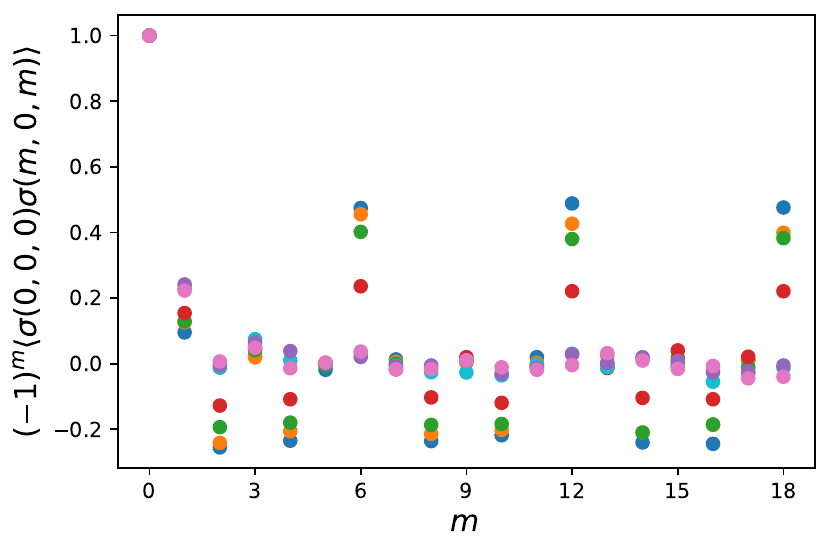}
    \caption{Spin correlations along the $\beta_{12}$ chain directions (left panel), and along $\beta_{13}$ (right panel), from Monte Carlo simulations of systems of size $L=9$ with $J_3'/J_1=0.8$ at temperatures near $T_c$ ($T_c/J_1 \approx 1.7$). The coordinate system has been chosen so that the selected chain pairs at low temperature are $\beta_{12}$ and $\beta_{34}$. Alignment of the vector spins along a chain corresponds to Ising variables that follow the pattern $+,-,+,-,...$; this results in positive values of $(-1)^{m}\langle \sigma(\mathbf{0}) \sigma(\mathbf{r}) \rangle$ for ${\bf r}$ along the $\beta_{12}$ or $\beta_{34}$ directions. The staggered behaviour of the correlators along other chain directions, an example of which is given in the right panel, is caused by the effective antiferromagnetic interaction between neighbouring $\beta_{12}$ or $\beta_{34}$ chains.}
    \label{fig:SScorrApp}
\end{figure}
\end{center}

The alignment of spins along a specific (spontaneously selected) pair of $\beta$-chain directions appears at the phase transition and grows rapidly upon lowering the temperature. This is most evident in the behaviour of $\langle \sigma(0) \sigma(\mathbf{r}) \rangle$ along the chains (see Fig.~\ref{fig:SScorrApp}). 
For $\mathbf{r}$ along the chosen chain direction $(-1)^{m}\langle \sigma(0) \sigma(\mathbf{r}) \rangle$ becomes positive and non-zero, consistent with a preferred alignment of spins head-to-tail along these chains. 

The behaviour of $(-1)^{m}\langle \sigma(0) \sigma(\mathbf{r}) \rangle$ for ${\bf r}$ along the other chain directions (the four not spontaneously selected for spin alignment), is more complex. Along these directions $(-1)^{m}\langle \sigma(0) \sigma(\mathbf{r}) \rangle$ is non-zero for even $m$. It is positive when $m$ is a multiple of 6, and otherwise negative (see lower panel of Fig.~\ref{fig:SScorrApp}). In this case, the spin-spin correlators are in fact probing correlations between the spontaneously selected $\beta$ chains. Consider, for example, the case where the selected chain pair is $\beta_{12}$ and $\beta_{34}$. If we start with a spin belonging to a $\beta_{12}$ chain (i.e., a spin on sublattice 1 or 2) and take an odd number of steps along any of the four directions defined by the remaining $\beta$ chains, we find a spin on a $\beta_{34}$ chain (i.e., a spin on sublattice 3 or 4). If we instead take an even number of steps we find a spin on a different $\beta_{12}$ chain. In App.~\ref{app:GSbuilding} we have shown that chains of the same type interact antiferromagnetically in the nematic phase, and that they form triangular patterns. The spin-spin correlations observed for $\mathbf{r}$ along chains not selected are in fact the correlations expected for a triangular lattice Ising antiferromagnet \cite{stephenson1964}. In this case that translates to positive correlation between chains of the same type and on the same triangular pattern sublattice (i.e., chains separated by $m$ a multiple of $6$) and negative correlation between chains of the same type but on different triangular pattern sublattices (i.e., other even values of $m$). 

There is no clear indication that the spin correlation grows as $T$ approaches $T_c$ from above, which is consistent with a first order phase transition. 
%
%
%%%
\onecolumngrid
\twocolumngrid
%%%
%
%

\begin{center}
\begin{figure}
    \centering
    \includegraphics[width=0.4\columnwidth]{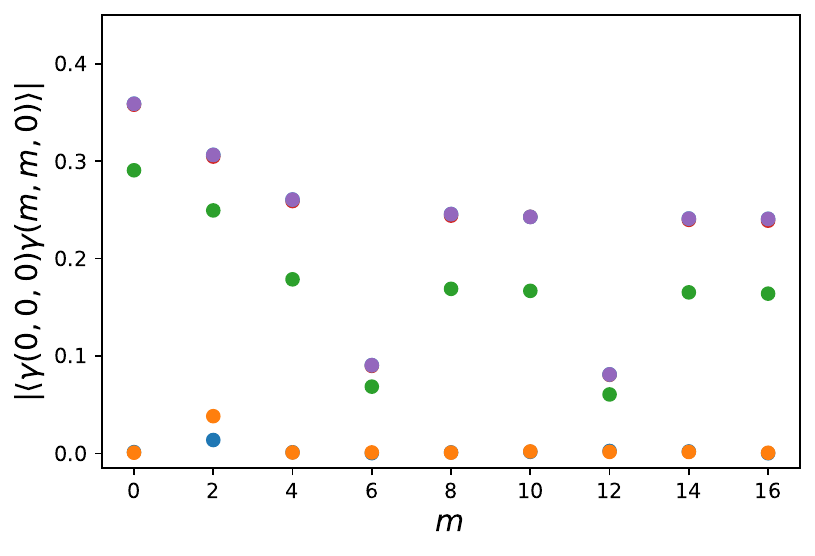}
    \includegraphics[width=0.4\columnwidth]{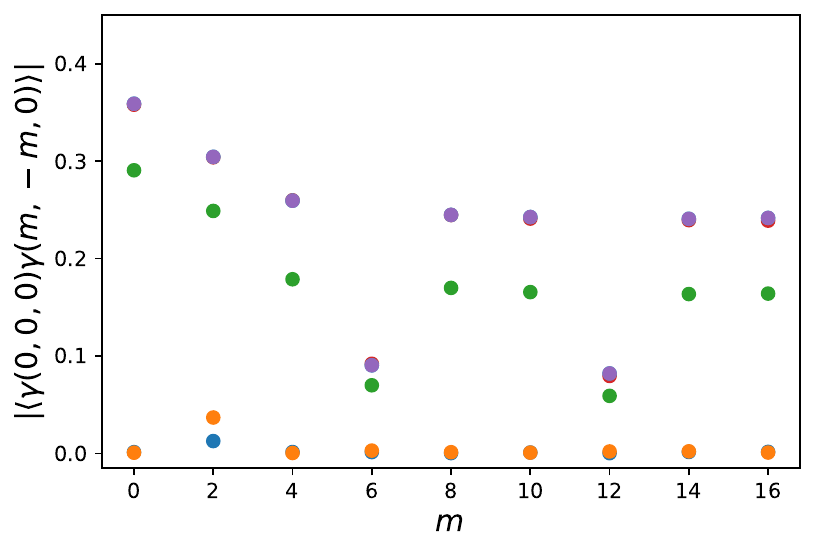}
    \includegraphics[width=0.4\columnwidth]{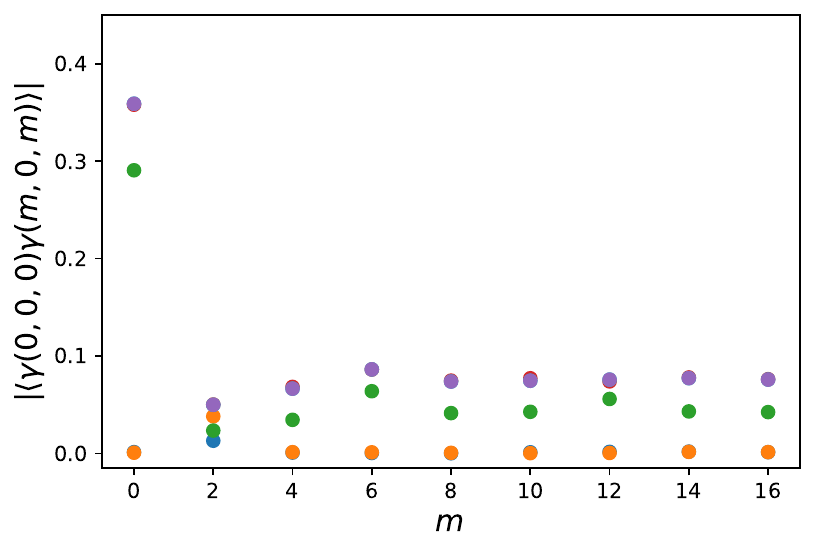}
    \includegraphics[width=0.4\columnwidth]{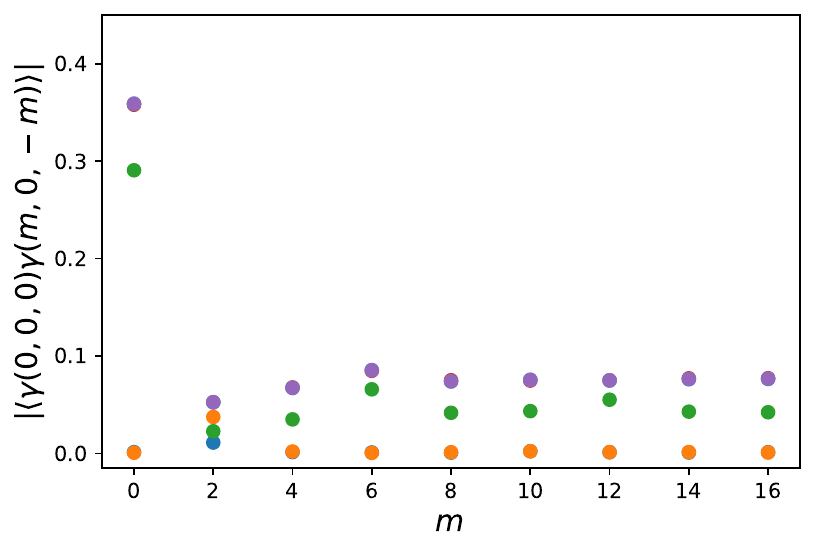}
    \includegraphics[width=0.4\columnwidth]{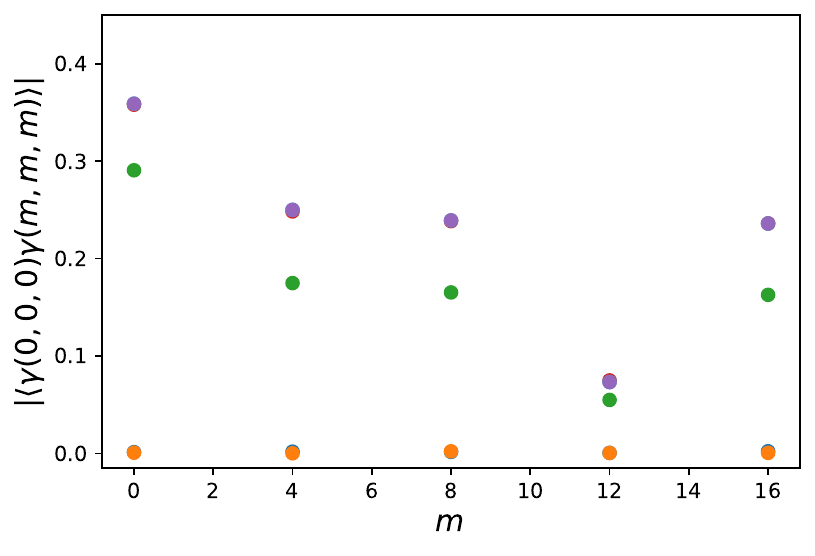}
    \includegraphics[width=0.4\columnwidth]{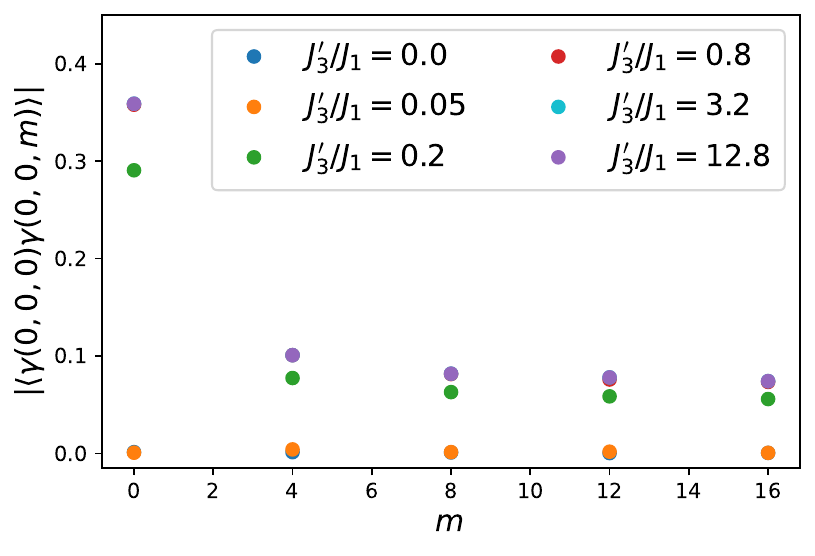}
    \caption{Correlations of the Potts variable $\gamma$, defined in Eq.~\eqref{eq:PottsVar}, along various symmetry directions. All plots are for $T/J_1 = 0.4$ (the temperature at which we measured the magnetic noise in Sec.~\ref{sec:PSD}), which is greater than $T_c$ for $J_3'/J_1 < 0.1$. For $J_3'/J_1 \geq 0.8$ the data points overlap perfectly. The states have been chosen so that the selected chain pairs at low temperature are $\beta_{12}$ and $\beta_{34}$, explaining why correlations are stronger along the [110] and [1-10] directions. The numerical results show no difference between the [100], [010], and [001] directions, which is why only the first one is shown. Similarly, the [101] and [011] directions are equivalent. This is for system size $L=9$. The smaller values when $m$ is a multiple of 6 for certain directions is again a symptom of the antiferromagnetic interactions between chains of the same type and the triangular pattern that they form.}
    \label{fig:GGcorrsApp}
\end{figure}
\end{center}

The correlators $\langle \gamma({\bf r}) \gamma({\bf r}') \rangle$ provide further evidence of the long range correlations that appear in the nematic spin ice phase. Again, correlations are particularly strong for ${\bf r} - {\bf r}'$ along specific symmetry directions corresponding to the directions of the selected pair of $\beta$-chains. In Fig.~\ref{fig:GGcorrsApp} the magnitude of $\langle \gamma(\mathbf{0})\gamma(\mathbf{r}) \rangle$ is plotted for $\mathbf{r}$ along several directions. Note once again that $\langle \gamma(\mathbf{0})\gamma(\mathbf{r}) \rangle \approx 0$ for all ${\bf r}$, ${\bf r}'$ if $T>T_c$. 
%
%
%%%
\onecolumngrid
\twocolumngrid
%%%
%
%

As previously stated, the nematic order develops in one of three possible distinct directions. When averaging over many independent simulations this must be taken into account, or the averaging would cancel out the spontaneous rotational symmetry breaking. We have accounted for this by choosing the coordinate system so that the low temperature, nematically ordered states all have preferential alignment along the $\beta_{12}$ and $\beta_{34}$ chains.
%
%
%%%%%%%%%%%%%%%%%%%%%%%%%%%%%%%%%%%%%%%%%%%%%%%%%%%%%%%%%%%

\section{Supercooling and numerical limits \label{app:Supercooling}}
\begin{figure}
    \centering
    \includegraphics[width=\columnwidth]{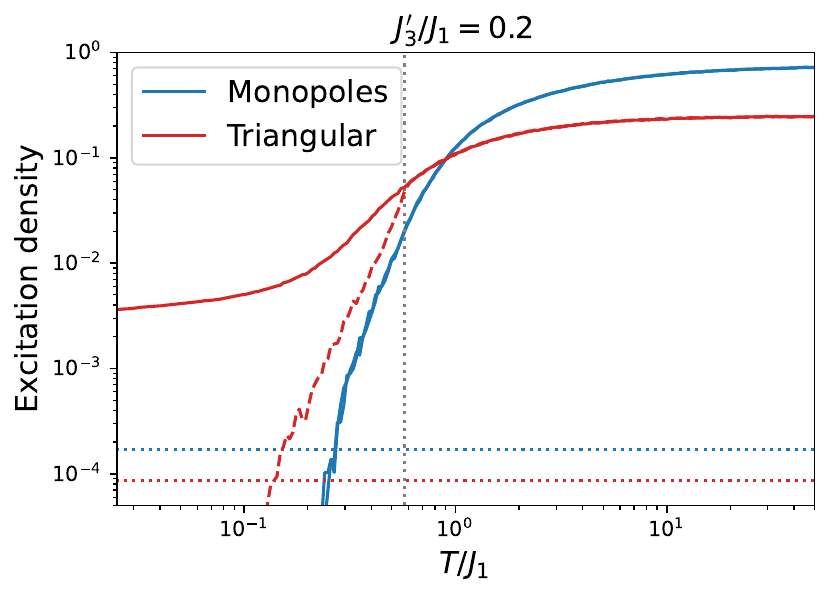}
    \includegraphics[width=\columnwidth]{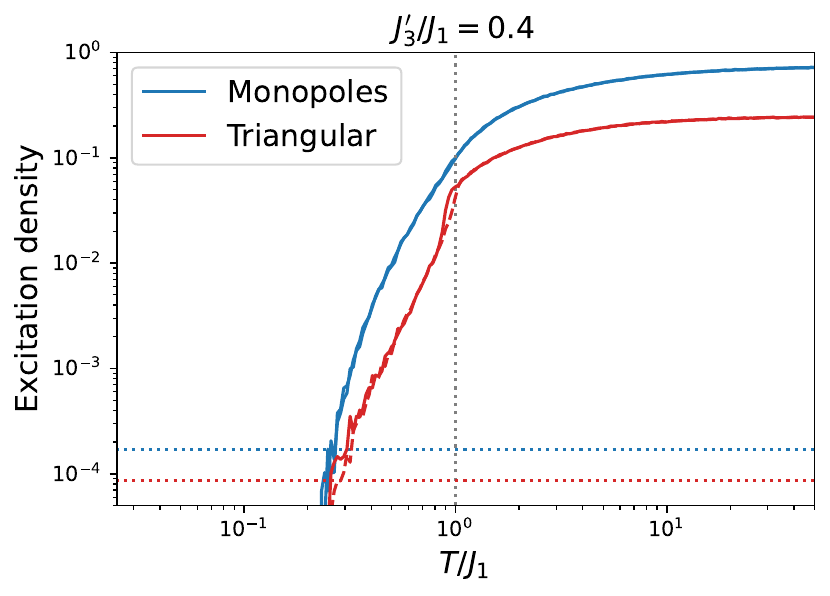}
    \caption{The densities of magnetic monopoles (blue) and triangular excitations (red) versus temperature for two values of $J_3'$. Cooling and heating runs are shown by solid and dashed lines, respectively. In the top panel, with $J_3'=0.2 J_1$, the system falls out of equilibrium when cooled below the equilibrium transition temperature, indicated by the dotted grey line. In the bottom panel, with $J_3'=0.4 J_1$, there is no evidence of low-temperature out-of-equilibrium behaviour (although some hysteresis is visible in the triangular excitation density). The dotted horizontal lines show the $1/V$ limit for the two types of excitations -- below these lines there is on average less than one excitation of the respective type in the entire finite system in our simulations. The figure is for $L=9$.}
    \label{fig:DensComp}
\end{figure}

In Fig.~\ref{fig:DensComp} we take a closer look at the excitation density curves for $J_3'=0.2 J_1$, where there are approximately $0.4$ monopoles per triangular excitation at the phase transition; and at $J_3'=0.4 J_1$, where there are approximately $1.9$ monopoles per triangular excitation at the phase transition. For $J_3'=0.2 J_1$ the triangular excitation densities measured in heating and cooling runs disagree below the phase transition, with the density plateauing for the cooling runs, thus indicating supercooled liquid behaviour. This is consistent with the idea that a minimum monopole density is required to enable the annihilation of triangular excitations. Based on these results we are tempted to conclude that a monopole density at least similar in magnitude to the triangular excitation density is required for the simulation to remain in equilibrium. 

Although the system becomes supercooled for $J_3' \lesssim 0.4J _1$, the triangular excitations are not completely frozen out: their density continues to decrease to a fairly small value, and $\Gamma$ indicates that some nematic order is present (Fig.~\ref{fig:orderpar}). In Sec.~\ref{sec:PSD} we find that the magnetisation noise becomes anomalous also in this supercooled regime.
%
%
%%%
%%%
%
%

\begin{figure}[h!]
    \centering
    \includegraphics[width=\columnwidth]{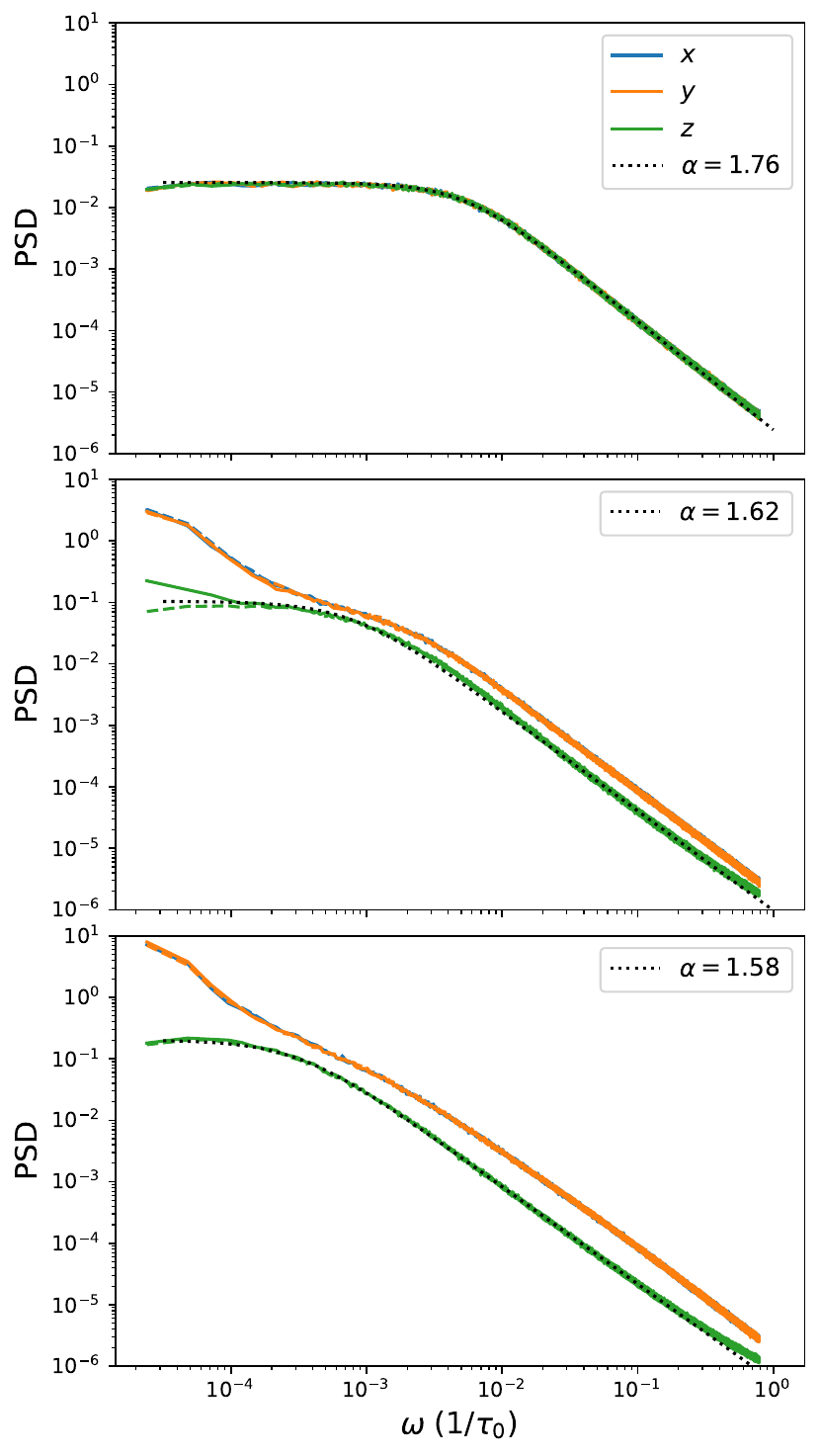}
    \caption{These figures show the PSD computed from the magnetisation measured along the three different Cartesian axes. They are all at $T/J_1=0.4$ and with $J_3'/J_1=0.1$ (top), $0.2$ (middle), and $0.8$ (bottom). Cooling runs are shown with solid lines and heating runs with dashed lines. The dotted black lines show the result of a fit with Eq.~\eqref{eq:fiteq}, with the extracted exponent shown in the legend. In the nematic spin ice phase, monopole transport along one axis (here chosen without loss of generality to be the $z$-axis) is reduced, as further discussed in Sec.~\ref{sec:Clusters} in the main text.}
    \label{fig:PSDs}
\end{figure}

An important observation is that even in the regime where the triangular excitation densities indicate that the system is supercooled, the PSDs computed from cooling and heating runs do not differ significantly (although some difference at low frequency is seen in the middle panel of Fig.~\ref{fig:PSDs}). Furthermore, the same type of anomalous noise is observed for larger $J_3'$ (see Fig.~\ref{fig:PSD_varyJ3p}), where there is no indication of supercooled behaviour. This demonstrates that supercooling is not the cause of the anomalous noise. %
%
%%%
%\onecolumngrid
%\begin{center}
%\end{center}
%\twocolumngrid
%%%
%
%

In the thermodynamic limit the excitation density strictly only goes to zero for $T=0$. In a finite system, however, the number of excitations can drop to zero at non-zero temperature. This $\sim 1/V$ threshold ($V$ being the finite size \quotes{volume} of the system) restricts the temperature regime one can generally study in simulations. For thermodynamic quantities one can arguably still obtain accurate results through extensive averaging. As the excitations are the drivers of the natural dynamics in the system, we ought to be weary of the $1/V$ limit when considering dynamical properties like the noise. 
Fig.~\ref{fig:DensComp} includes the $1/V$ thresholds for the two excitation types, and shows that the supercooled behaviour occurs well before either density drops close to this level.

%
%
%%%%%%%%%%%%%%%%%%%%%%%%%%%%%%%%%%%%%%%%%%%%%%%%%%%%%%%%%%%
\bibliography{references}

%%%%%%%%%%%%%%%%%%%%%%%%%%%%%%%%%%%%%%%%%%%%%%%%%%%%%%%%%%%
%%%%%%%%%%%%%%%%%%%%%%%%%%%%%%%%%%%%%%%%%%%%%%%%%%%%%%%%%%%
\end{document}